\documentclass[onecolumn,usenatbib]{mn2e}
\usepackage{amsmath}
\usepackage[dvips]{graphicx}
\begin{document}
\title{Relativistic Models of Galaxies}
\author[D. Vogt and P. S. Letelier]{D. Vogt$^1$\thanks{e-mail: danielvt@ifi.unicamp.br } and P. S. Letelier$^2$
\thanks{e-mail: letelier@ime.unicamp.br; URL: http://www.ime.unicamp.br/{\char126}letelier/}\\
$^1$Instituto de F\'{\i}sica Gleb Wataghin, Universidade Estadual de Campinas
13083-970 Campinas, S.\ P., Brazil \\
$^2$ Departamento de Matem\'{a}tica Aplicada-IMECC, Universidade Estadual
de Campinas 13083-970 Campinas, S.\ P., Brazil}
\maketitle
\begin{abstract}
A special form of the isotropic metric in cylindrical coordinates is used to construct what may 
be interpreted as the General Relativistic versions of some well known potential-density pairs used in Newtonian 
gravity to model three-dimensional distributions of matter in galaxies. The components of
the energy-momentum tensor are calculated for the first two 
Miyamoto-Nagai potentials and a particular potential due to Satoh. The three potentials yield 
distributions of matter in which all tensions are pressures and all energy conditions are satisfied for 
certain ranges of the free parameters. A few 
non-planar geodesic orbits are computed for one of the potentials and compared with the Newtonian 
case. Rotation is also incorporated to the models and the effects of the source rotation on the rotation profile are calculated as first order corrections by 
using an approximate form of the Kerr metric in isotropic coordinates.
\end{abstract}
\section{Introduction}

Analytical axially symmetric solutions of gravitating matter provide very accurate 
models for the mass distribution of galaxies. In Newtonian theory, models for
globular clusters and spherical galaxies were presented by \citet{b1} 
and \citet{b2}. On the other hand, to describe the mass distribution within 
highly flattened axisymmetric galaxies, \citet{b3} found a family 
of density potential pairs, the first of them was first derived by \citet{b4}. 
Later, Miyamoto and Nagai (\citealt{b5}; \citealt{b6}) ``thickened-up'' Toomre's series
of disk models to yield pairs of three-dimensional potential and density functions. In a 
similar way, \citet{b7} obtained a family of three-dimensional axisymmetric 
mass distributions from the higher order Plummer models. For a detailed account and 
the description of other potential-density pairs used in galactic models, see for example \citet{b8}.

In General Relativity, several exact disk-like solutions have been found, among them the 
static disks without radial pressure studied by \citet{b9} and \citet{b10}, and disks with radial pressure by 
 \citet{b11}. 
Several classes of exact solutions
of the Einstein field equations corresponding to static thin disks with or without radial 
pressure have been obtained by different authors (\citealt{b12}; \citealt{b13}; \citealt{b14}; 
\citealt{b15}; \citealt*{b16}; \citealt*{b17}; \citealt{b18}; \citealt{b19}; \citealt{b20}; 
\citealt{b21}). 
Thin rotating disks that can be considered as a source of the Kerr metric were presented 
by \citet{b22}, while rotating disks with heat flow were studied by \citet{b23}. 
Also thin disks with radial tension \citep{b24}, magnetic fields \citep{b25} 
and magnetic and electric fields \citep*{b26} were considered. The nonlinear superposition 
of a disk and a black hole was first obtained by  \citet{b15}. Perfect
fluid disks with halos were studied by \citet{b27} and charged perfect fluid disks by \citet{b28}. 
The stability of some General Relativistic thin disk models using a first order perturbation of
the energy-momentum tensor was investigated by \citet{b29}. For a survey on self gravitating relativistic
disks, see for instance \citet{b30}.

In the works cited above an inverse style method was used to solve the Einstein equations, 
i.\ e., the energy-momentum tensor is computed from the metric representing the disk. Another 
approach to generate disks is by solving the Einstein equations given a source (energy-momentum
tensor). Essentialy, they are obtained by solving a Riemann-Hilbert problem and are highly nontrivial. This
has been used by the Jena group to generate several exact solutions of disks (\citealt{b31}; 
\citealt{b32}; \citealt{b33}; \citealt{b34}; \citealt{b35}; \citealt{b36}; \citealt{b37}; 
\citealt{b38}). 

Recently \citet{b39} constructed models of static relativistic thick disks in
various coordinate systems. Although the disks have constant thickness, the matter density decreases
rapidly with radius and the $z$ coordinate, and in principle they also may be used to represent both the disk 
part and the central bulges of galaxies.

The aim of this work is to consider more realistic three-dimensional models
for the gravitational field of galaxies in a General Relativistic context. The distribution of matter 
fills all the space, but the total mass is finite. Essentially we formulate General Relativistic
versions in isotropic coordinates of the potential-density pairs deduced by Miyamoyo and 
Nagai (\citealt{b5}; \citealt{b6}) and by \citet{b7}. Also the influence of  galactic rotation 
on the rotation profile of test particles on the galactic plane is estimated using an 
approximate form of the Kerr metric in isotropic coordinates. 

The work is divided as follows. In Sec.\ \ref{sec_isotropic} we present a special form 
of the isotropic metric and the components of the energy-momentum tensor as functions of 
the metric coefficients, from which the physical properties of the galactic matter distributions 
can be calculated. We also derive expressions for the tangential circular velocity 
and specific angular momentum of test particles in circular motion on the 
galactic plane. These results are then applied in Sec.\ \ref{sec_miy} and Sec.\ 
\ref{sec_satoh} to construct General Relativistic analogues of two Miyamoto-Nagai 
models and one galactic model proposed by Satoh. As an example of application of these 
models, in Sec.\ \ref{sec_geo} we calculate a few geodesic orbits and compare them 
with the Newtonian case. In Sec.\ \ref{sec_rot} we deduce the first 
order effects of galactic rotation on the tangential circular velocity of
geodesic motion of particles on the galactic plane using an
approximate form of the Kerr metric in isotropic coordinates, and apply the result on the 
above mentioned models. Finally, we summarize our results in Sec.\ \ref{sec_discuss}.

\section{Einstein Equations in Isotropic Coordinates} \label{sec_isotropic}

Let us consider the particular case of axial symmetric space-times  whose metric
can be writtem in an isotropic form  in cylindrical coordinates
$(t,R,z,\varphi)$, 
\begin{equation} \label{eq_line1}
\mathrm{d}s^2=e^{\nu(R,z)}c^2\mathrm{d}t^2-e^{\lambda(R,z)}\left( \mathrm{d}R^2+
\mathrm{d}z^2+R^2\mathrm{d}\varphi ^2 \right) \mbox{.}
\end{equation}
In order to construct a General Relativistic formulation of the known
Newtonian models for the mass distribution of galaxies, it is
convenient to use a particular form of metric Eq.\ (\ref{eq_line1}), suggested by generalising the form
of Schwarzschild's metric in isotropic coordinates
\begin{equation} \label{eq_metr_iso}
\mathrm{d}s^2=\left( \frac{1-f}{1+f}\right)^2 c^2\mathrm{d}t^2-(1+f)^4\left( \mathrm{d}R^2+
\mathrm{d}z^2+R^2\mathrm{d}\varphi ^2 \right) \mbox{,}
\end{equation}
where $f=f(R,z)$. Metric Eq.\ (\ref{eq_metr_iso}) gives the Schwarzschild solution 
when $f$ is taken as 
\begin{equation} \label{eq_iso_schw}
f=\frac{GM}{2c^2\sqrt{R^2+z^2}} \mbox{.}
\end{equation}
For metric Eq.\ (\ref{eq_metr_iso}), the Einstein equations $G_{\mu \nu}=-\frac{8 \pi G}{c^4}T_{\mu \nu}$
yield the following expressions for the components of the energy-momentum
tensor\footnote{The Einstein tensor was computed using MAPLE and the result was checked using REDUCE.}
\begin{subequations}
\begin{align}
T^t_t &=-\frac{c^4}{2\pi G(1+f)^5} \left( f_{,RR}+f_{,zz}+\frac{f_{,R}}{R} \right) \mbox{,} \label{eq_Ttt} \\
T^R_R &=\frac{c^4}{4\pi G (1+f)^5(1-f)} \left( ff_{,zz}+\frac{ff_{,R}}{R}+2f_{,R}^2-f_{,z}^2 \right) \mbox{,} \label{eq_TRR} \\
T^z_z &=\frac{c^4}{4\pi G (1+f)^5(1-f)} \left( ff_{,RR}+\frac{ff_{,R}}{R}+2f_{,z}^2-f_{,R}^2 \right) \mbox{,} \label{eq_Tzz} \\
T^R_z &=T^z_R= -\frac{c^4}{4\pi G (1+f)^5(1-f)} \left( ff_{,Rz}-3f_{,R}f_{,z} \right) \mbox{,} \label{eq_TRz} \\
T^{\varphi}_{\varphi} &= \frac{c^4}{4\pi G (1+f)^5(1-f)} \left[ f \left( f_{,RR}+f_{,zz} \right) -f_{,R}^2-f_{,z}^2 \right]
\mbox{.} \label{eq_Tfifi}
\end{align}
\end{subequations}
The energy density is given by $\sigma=T^t_t/c^2$, and the stresses (pressures or tensions) along a
particular direction read $P_i=-T^i_i$ when the energy-momentum tensor is
diagonal. It is a surprising fact that the  component $T^t_t$ is proporcional
to the usual Laplacian of the function $f$   in $flat$ cylindrical coordinates.
Note that in the Newtonian limit when $f \ll 1$, Eq.\ 
(\ref{eq_Ttt}) reduces to Poisson equation
\begin{equation} \label{eq_poisson}
\nabla^2 \Phi = 4\pi G \rho_N \mbox{,}
\end{equation}
if the function $f$ is related to the gravitational potential $\Phi$ by
\begin{equation} \label{eq_f}
f=-\frac{\Phi}{2c^2} \mbox{.}
\end{equation}

The energy-momentum tensor will be diagonal $(T^R_z=T^z_R=0)$ provided $f$ 
has the form
\begin{equation} \label{eq_f_best}
f=\frac{C}{\sqrt{w(R)+g(z)}} \mbox{,}
\end{equation}
where $C$ is a constant and $w(R)$ and $g(z)$ are arbitrary functions. Furthermore, the
components $T^R_R$ and $T^{\varphi}_{\varphi}$ will be equal (isotropic radial and
azimuthal stresses) only if $w(R)=R^2$.

Two physical quantities of interest related to the circular motion of
test particles along geodesics on the galactic plane are the tangential velocity $v_c$ (rotation profile)
and the specific angular momentum $h$. The assumption of geodesic motion is valid for the case of
 a particle moving in a very  diluted  gas like the gas made of stars that models a galaxy disk.
For metric Eq.\ (\ref{eq_line1}) the expressions are \citep{b27} 
\begin{align}
v_{c}^2 &=c^2R^2\frac{e^{\lambda}(e^{\nu})_{,R}}{e^{\nu}(R^2e^{\lambda})_{,R}} \mbox{,} \label{eq_vc} \\
h &=cR^2e^{\lambda}\sqrt{\frac{(e^{\nu})_{,R}}{e^{\nu}(R^2e^{\lambda})_{,R}-R^2e^{\lambda}(e^{\nu})_{,R}}}
\mbox{,} \label{eq_h}
\end{align}
where all functions are evaluated on $z=0$. The angular momentum can be used to determine the stability of circular orbits on the galactic plane by using an extension
of the Rayleigh criteria of stability of a fluid at rest in a gravitational field (\citealt{b40}, see also \citealt{b41})
\begin{equation} \label{eq_stab}
\left. h\frac{dh}{dR} \right|_{z=0}>0 \mbox{.}
\end{equation}
For the specific metric form Eq.\ (\ref{eq_metr_iso}), Eq.\ (\ref{eq_vc})--(\ref{eq_h}) simplify to
\begin{align}
v_c^2 &=-\frac{2c^2Rf_{,R}}{( 1-f )( 1+f+2Rf_{,R})} \mbox{,} \label{eq_vcf} \\
h &=cR^2\left( 1+f \right)^2 \sqrt{\frac{-2f_{,R}}{R\left[1-f^2+2Rf_{,R}(2-f) \right]}} \mbox{,} \label{eq_hf}
\end{align}
\section{General Relativistic Miyamoto-Nagai Models} \label{sec_miy}

\subsection{First Model} \label{sub_sec_miy1}

The simplest gravitational potential proposed by Miyamoto and Nagai (\citealt{b5}; \citealt{b6}) to
represent stratifications of mass in the central bulges and in the disk parts 
of galaxies
 is given by
\begin{equation}
\Phi(R,z)=-\frac{GM}{\sqrt{R^2+(a+\sqrt{z^2+b^2})^2}} \mbox{,}
\end{equation}
where $a,b$ are positive constants. The corresponding three-dimensional 
density distribution $\rho_N(R,z)$ is easily derived from Poisson equation (\ref{eq_poisson})
\begin{equation} \label{eq_newt_miy1}
\rho_N(R,z)=\frac{b^2M}{4\pi} \frac{aR^2+(a+3\sqrt{z^2+b^2})(a+\sqrt{z^2+b^2})^2}
{\left[ R^2+(a+\sqrt{z^2+b^2})^2\right]^{5/2}(z^2+b^2)^{3/2}} \mbox{.}
\end{equation}
Now we consider the function $f(R,z)$ defined in the previous section as
\begin{equation} \label{eq_f_miyam1}
f(R,z)=\frac{GM}{2c^2\sqrt{R^2+(a+\sqrt{z^2+b^2})^2}} \mbox{.}
\end{equation}
Substituting Eq.\ (\ref{eq_f_miyam1}) in Eq.\ (\ref{eq_Ttt})--(\ref{eq_Tfifi}) we find the following 
expressions for the nonzero components of the energy-momentum tensor
\begin{subequations}
\begin{align}
\bar{\sigma} &=\frac{\tilde{b}^2 \left[ \tilde{a}\tilde{R}^2+(\tilde{a}+\xi)^2(\tilde{a}+3 \xi) \right]}
{4 \pi \xi ^3 \left[1+\sqrt{\tilde{R}^2+(\tilde{a}+\xi)^2} \right]^5} \mbox{,} \label{eq_sigma_miy1}\\
\bar{p}_R &=\bar{p}_{\varphi}=\frac{\tilde{b}^2}{8 \pi \xi ^3}
\frac{\left[ \tilde{a}\tilde{R}^2+(\tilde{a}+\xi)^2(\tilde{a}+2 \xi) \right]}
{\left[ 1+\sqrt{\tilde{R}^2+(\tilde{a}+\xi )^2} \right]^5\left[ -1+\sqrt{\tilde{R}^2+(\tilde{a}+\xi )^2} \right]}
\mbox{,} \label{eq_pr_miy1}\\
\bar{p}_z &=\frac{\tilde{b}^2(\tilde{a}+\xi )^2}
{4 \pi \xi ^2 \left[ 1+\sqrt{\tilde{R}^2+(\tilde{a}+\xi )^2} \right]^5\left[ -1+\sqrt{\tilde{R}^2+(\tilde{a}+\xi )^2} \right]}
\mbox{,} \label{eq_pz_miy1}
\end{align}
\end{subequations}
where the variables and parameters were rescaled in terms of the Schwarzschild radius in isotropic spherical
coordinates  $r_{s}=GM/(2c^2)$\footnote{Note that the Schwarzschild radius $r_s$ in isotropic spherical 
coordinates is related to Schwarzschild radius $r'_s$ in canonical spherical coordinates by $r_s=r'_s/4$.}: 
$\tilde{R}=R/r_s$, $\tilde{z}=z/r_s$, $\tilde{a}=a/r_s$, $\tilde{b}=b/r_s$,
$\sigma=\frac{M}{r_s^3}\bar{\sigma}$, $p_i=\frac{Mc^2}{r_s^3}\bar{p}_i$ and $\xi=\sqrt{\tilde{z}^2+\tilde{b}^2}$.
The ``effective Newtonian density'' $\rho=\sigma+p_R/c^2+p_{\varphi}/c^2+p_z/c^2$ reads
\begin{equation} \label{eq_rho_miy1}
\bar{\rho}=\frac{\tilde{b}^2\sqrt{\tilde{R}^2+(\tilde{a}+\xi )^2} \left[ \tilde{a}\tilde{R}^2+
(\tilde{a}+\xi )^2(\tilde{a}+3 \xi ) \right]}
{4 \pi \xi ^3\left[ 1+\sqrt{\tilde{R}^2+(\tilde{a}+\xi )^2} \right]^5\left[ -1+\sqrt{\tilde{R}^2+(\tilde{a}+\xi )^2} \right]}
\mbox{,}
\end{equation}
with $\rho=\frac{M}{r_s^3}\bar{\rho}$. 

The strong energy condition requires that $\rho \geq 0$, whereas the weak energy condition imposes the
condition $\sigma \geq 0$. The dominant energy condition requires $|p_R/\sigma| \leq c^2$, $|p_z/\sigma| \leq c^2$ and
$|p_{\varphi}/\sigma| \leq c^2$. Eq.\ (\ref{eq_sigma_miy1})--(\ref{eq_rho_miy1}) show that the densities 
are positive everywhere and so are the stresses (pressures) if we impose the condition $\tilde{a}+\tilde{b}>1$.
Moreover, we have isotropic pressures along the radial and azimuthal directions. When $\tilde{R}$, $\tilde{z}$, 
$\tilde{a}$, $\tilde{b} \gg 1$, the first order expansion of Eq.\ (\ref{eq_rho_miy1}) is
\begin{equation}
\bar{\rho} \approx \frac{\tilde{b}^2\left[ \tilde{a}\tilde{R}^2+
(\tilde{a}+\xi )^2(\tilde{a}+3 \xi ) \right]}{4 \pi \xi^3\left[ \tilde{R}^2+(\tilde{a}+\xi )^2 \right]^{5/2}}
\left[ 1-\frac{4}{\sqrt{\tilde{R}^2+(\tilde{a}+\xi )^2}} \right] \mbox{.}
\end{equation}
Thus the first order General Relativistic correction lowers the mass density.

\begin{figure}
\centering
\includegraphics[scale=0.5]{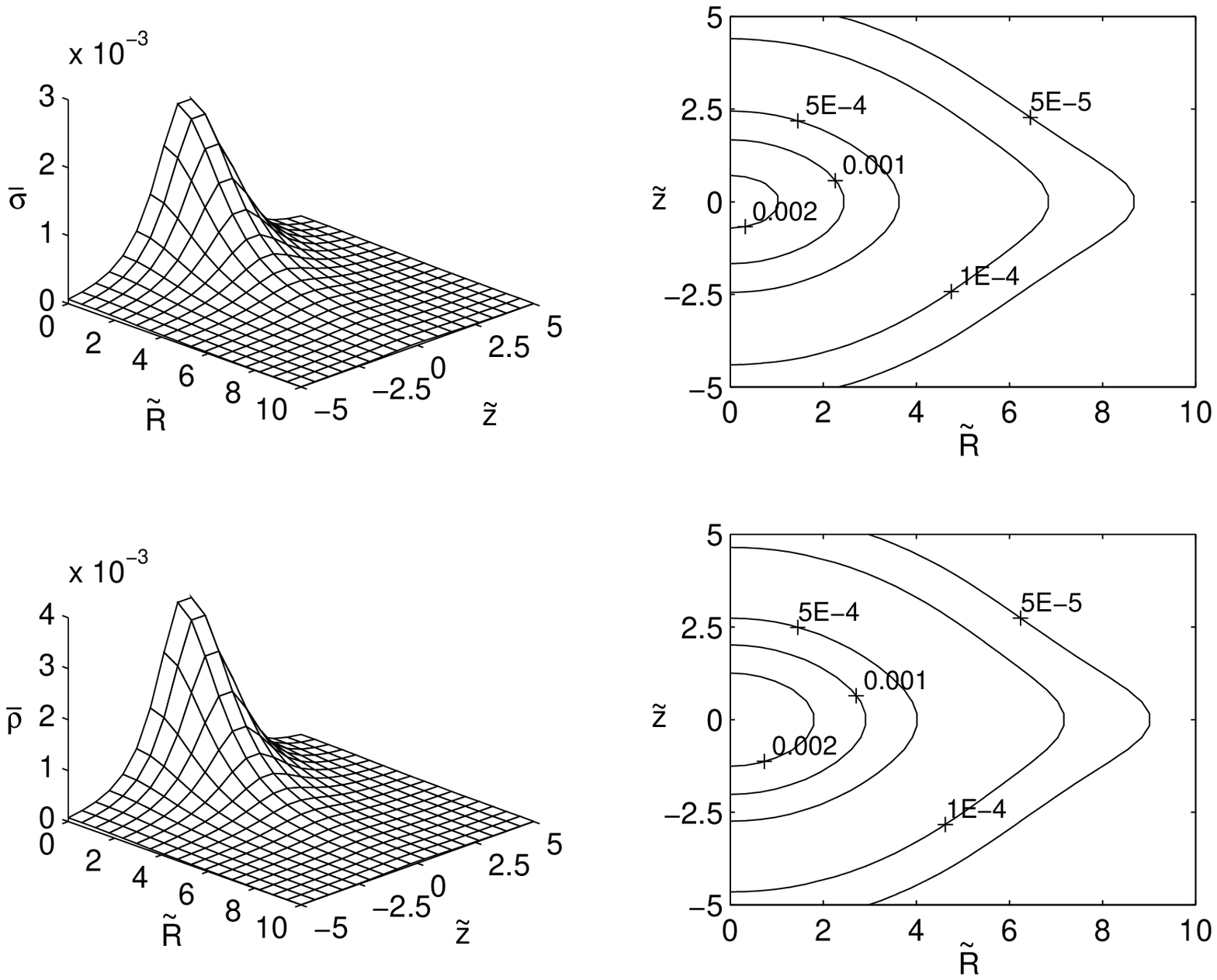}
\caption{Surface plot and level curves of the energy density $\bar{\sigma}$ Eq.\ (\ref{eq_sigma_miy1}) and ``effective
Newtonian density'' $\bar{\rho}$ Eq.\ (\ref{eq_rho_miy1}) with parameters $\tilde{a}=1$ and $\tilde{b}=2$.} \label{fig1}
\end{figure}

\begin{figure}
\centering
\includegraphics[scale=0.5]{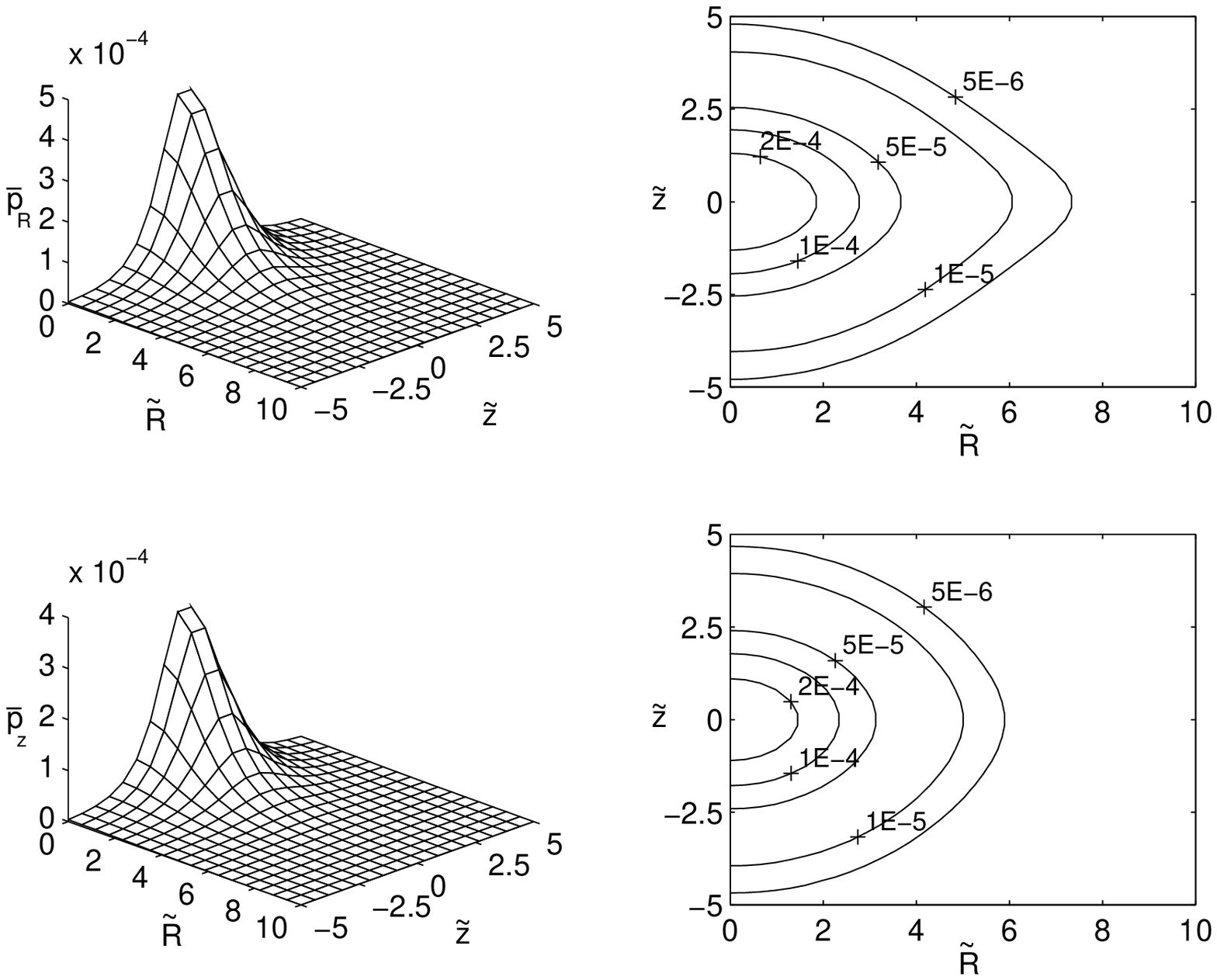}
\caption{Surface plot and level curves of the radial pressure $\bar{p}_{R}$ Eq.\ (\ref{eq_pr_miy1}) and vertical
pressure $\bar{p}_{z}$ Eq.\ (\ref{eq_pz_miy1}) with parameters $\tilde{a}=1$ and $\tilde{b}=2$.} \label{fig2}
\end{figure}

\begin{figure}
\centering
\includegraphics[scale=0.5]{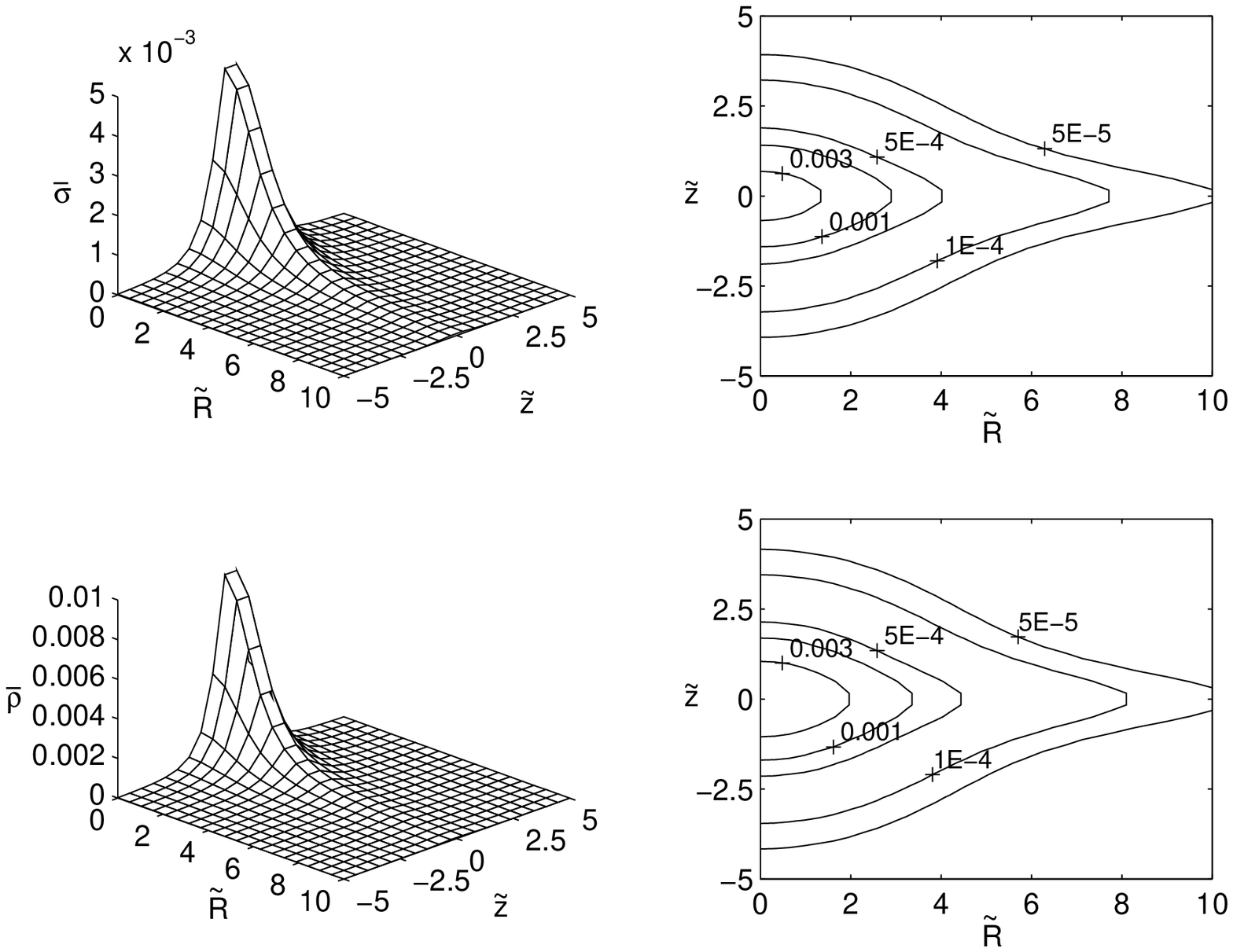}
\caption{Surface plot and level curves of the energy density $\bar{\sigma}$ Eq.\ (\ref{eq_sigma_miy1}) and ``effective
Newtonian density'' $\bar{\rho}$ Eq.\ (\ref{eq_rho_miy1}) with parameters $\tilde{a}=1$ and $\tilde{b}=1$.} \label{fig3}
\end{figure}

\begin{figure}
\centering
\includegraphics[scale=0.5]{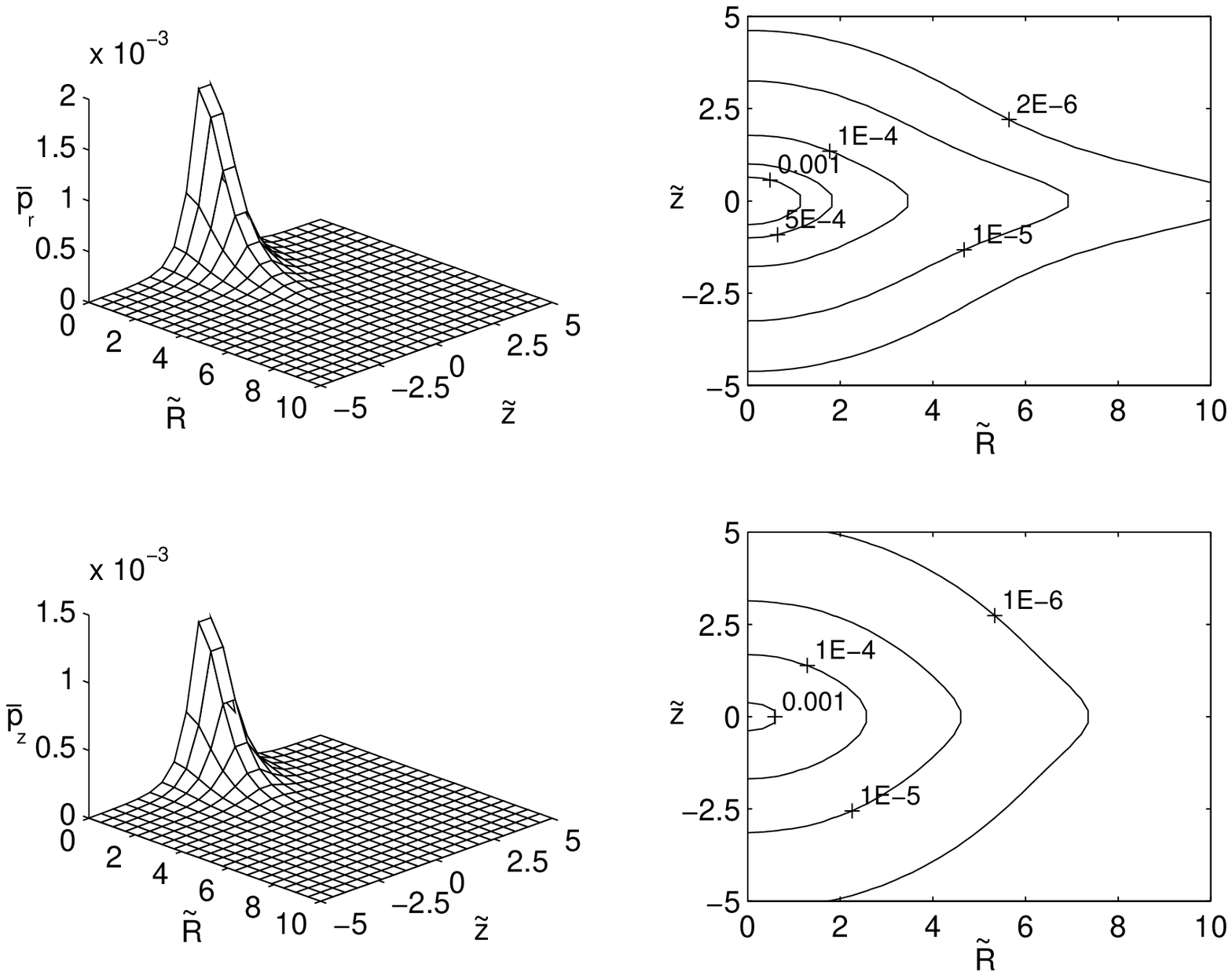}
\caption{Surface plot and level curves of the radial pressure $\bar{p}_{R}$ Eq.\ (\ref{eq_pr_miy1}) and vertical
pressure $\bar{p}_{z}$ Eq.\ (\ref{eq_pz_miy1}) with parameters $\tilde{a}=1$ and $\tilde{b}=1$.} \label{fig4}
\end{figure}

In Fig.\ \ref{fig1} and Fig.\ \ref{fig2} we show the surfaces and level curves of the densities and pressures
Eq.\ (\ref{eq_sigma_miy1})--(\ref{eq_rho_miy1}) for parameters $\tilde{a}=1$, $\tilde{b}=2$ and the
same physical quantities for parameters $\tilde{a}=1$, $\tilde{b}=1$ are displayed in Fig.\ \ref{fig3} -- \ref{fig4}.
As in the Newtonian case, the less the ratio $b/a$, the flatter are the mass distributions and the pressure 
distributions show similar behaviour. For $\tilde{a}=1$, $\tilde{b}=2$ we have $\tilde{p}_R/\tilde{\sigma}<0.2$ and
$\tilde{p}_z/\tilde{\sigma}<0.15$ and for $\tilde{a}=1$, $\tilde{b}=1$ the ratios are $\tilde{p}_R/\tilde{\sigma}<0.4$ and
$\tilde{p}_z/\tilde{\sigma}<0.3$. Thus for these parameters all energy conditions are satisfied.

The tangential circular velocity Eq.\ (\ref{eq_vcf}) and specific angular momentum Eq.\ (\ref{eq_hf}) for
$f$ given by Eq.\ (\ref{eq_f_miyam1}) evaluated on $z=0$ read
\begin{align}
& \left( \frac{v_c}{c} \right)^2 =\frac{2\tilde{R}^2\sqrt{\tilde{R}^2+\eta^2}}
{\left( -1+\sqrt{\tilde{R}^2+\eta^2}\right)
\left[ -\tilde{R}^2+\eta^2+(\tilde{R}^2+\eta^2)^{3/2}\right]} \mbox{,} \label{eq_vc_miy1} \\
& \bar{h} =\frac{\sqrt{2}\tilde{R}^2\left( 1+\sqrt{\tilde{R}^2+\eta^2}\right)^2}{(\tilde{R}^2+\eta^2)^{3/4}
\sqrt{ (\tilde{R}^2+\eta^2)^2 +\tilde{R}^2-\eta^2-4\tilde{R}^2\sqrt{\tilde{R}^2+\eta^2}}} \mbox{,} \label{eq_h_miy1}
\end{align}
where we defined $h=cr_s\bar{h}$ and $\eta=\tilde{a}+\tilde{b}$. The first
order expansion of Eq.\ (\ref{eq_vc_miy1}) reads
\begin{equation} \label{eq_vc_miy1c}
\frac{v_c}{c}=\frac{\tilde{R}\sqrt{2}}{\left[ \tilde{R}^2+(\tilde{a}+\tilde{b})^2 \right]^{3/4}} \left\{
1+\frac{\tilde{R}^2}{\left[ \tilde{R}^2+(\tilde{a}+\tilde{b})^2 \right]^{3/2}} \right\} \mbox{.}
\end{equation}

\begin{figure}
\centering
\includegraphics[scale=0.68]{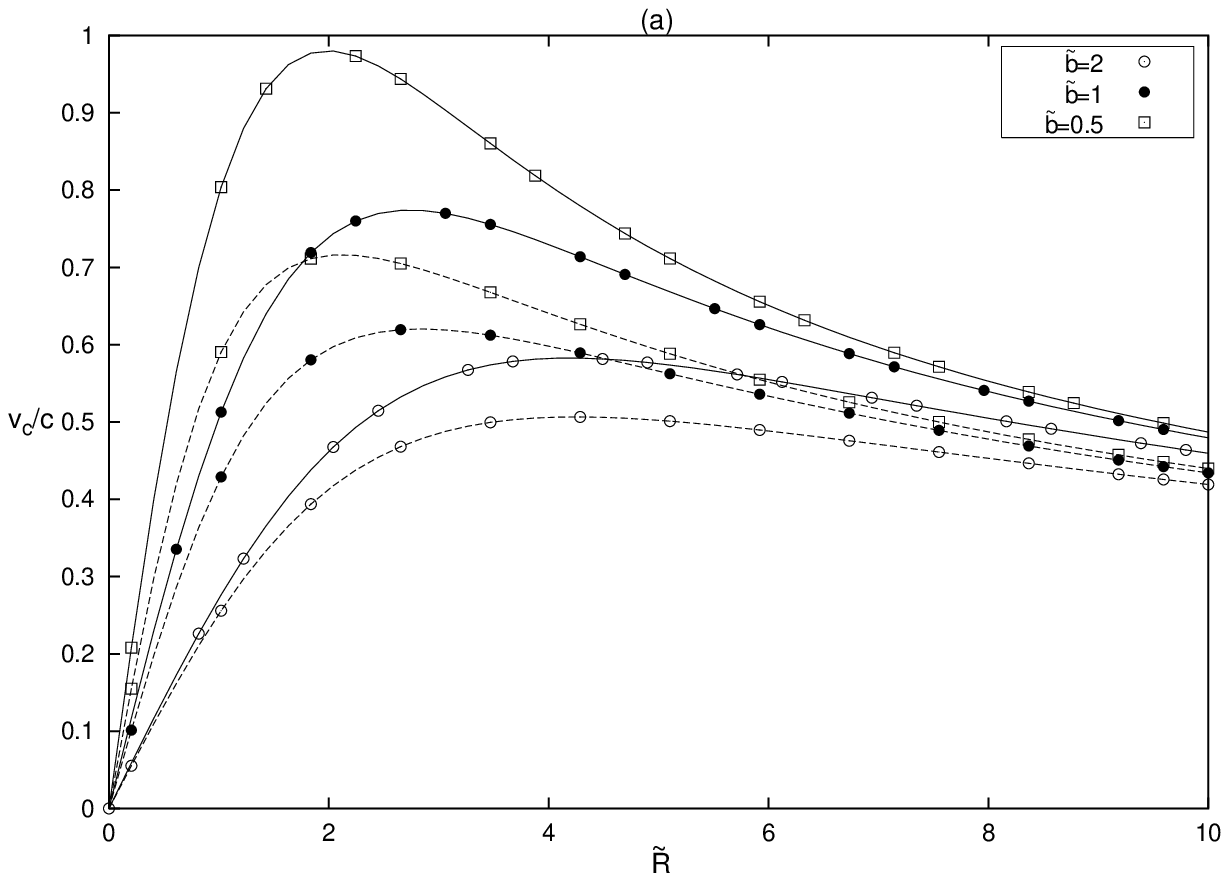}\\
\vspace{0.2cm}
\includegraphics[scale=0.68]{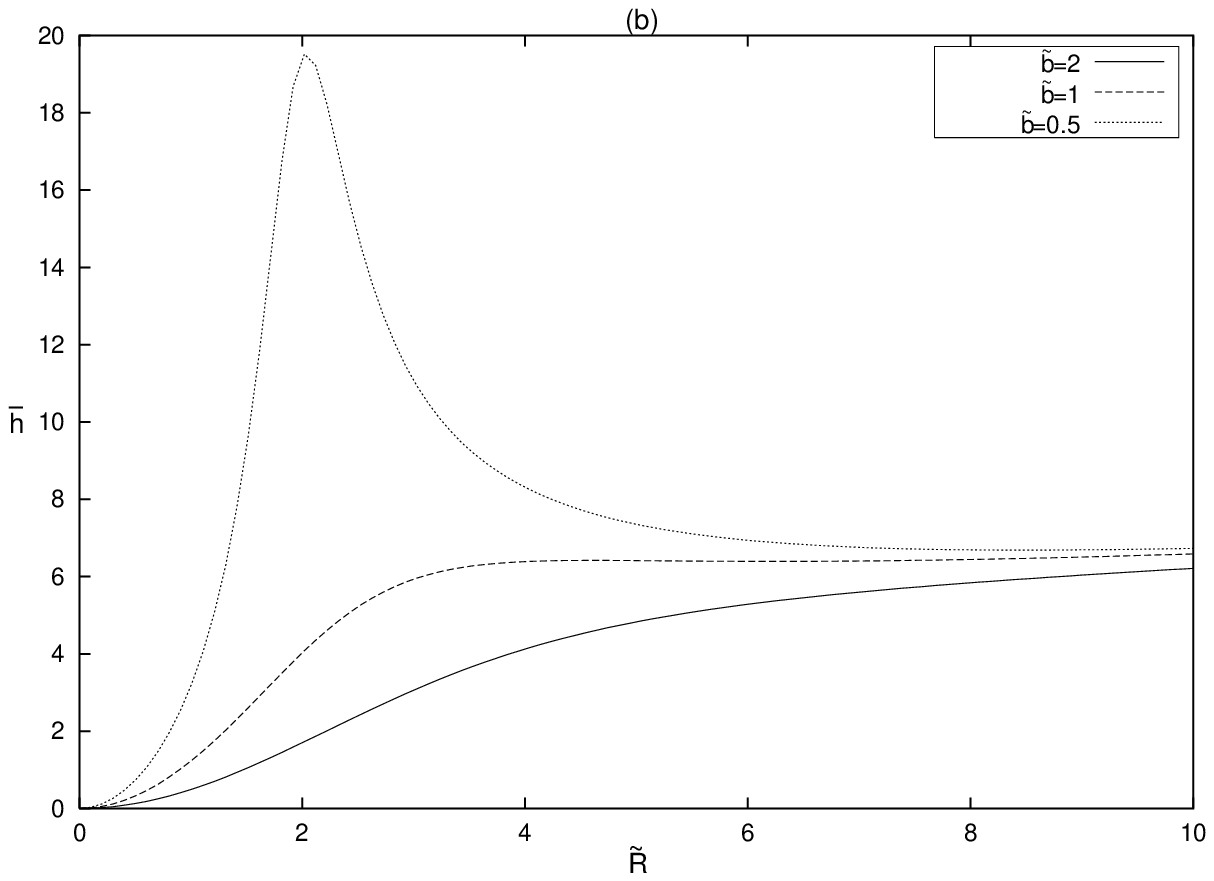}
\caption{(a) Relativistic rotation profile Eq.\ (\ref{eq_vc_miy1}) and Newtonian rotation profile
(dashed curves) with parameters $\tilde{a}=1$ and $\tilde{b}=2$, $\tilde{b}=1$ and $\tilde{b}=0.5$. (b)
Specific angular momentum Eq.\ (\ref{eq_h_miy1}) for the same parameters.} \label{fig5}
\end{figure}

Fig.\ \ref{fig5} (a)--(b) shows, respectively, the rotation profile Eq.\ (\ref{eq_vc_miy1}) and the curves of
specific angular momentum Eq.\ (\ref{eq_h_miy1}) for parameters $\tilde{a}=1$ and $\tilde{b}=2$, $\tilde{b}=1$
and $\tilde{b}=0.5$. In Fig.\ \ref{fig5} (a) the dashed curves represent the newtonian tangential velocities.
Relativistic tangential velocities are always higher than the Newtonian ones, as suggested by the first
order correction in Eq.\ (\ref{eq_vc_miy1c}). The curves of specific angular momentum show that
circular orbits on the plane $\tilde{z}=0$ turn to be unstable for higher flattened matter distributions (the curve for $\tilde{b}=0.5$).
The orbits for the parameters $\tilde{a}=1$, $\tilde{b}=2$ are all stable, whereas in the case $\tilde{b}=1$ 
a small region of instability appears between $\tilde{R} \approx 5$ and $\tilde{R} \approx 6$. The Newtonian limit of Eq.\ (\ref{eq_h_miy1}) gives
\begin{equation}
\bar{h}=\frac{\sqrt{2}\tilde{R}^2}{[\tilde{R}^2+(\tilde{a}+\tilde{b})^2]^{3/4}} \mbox{,}
\end{equation}
from which is easily checked that $\frac{d\bar{h}}{d\tilde{R}} \geq 0$ for all $\tilde{a}$, $\tilde{b}$ 
and $\tilde{R}$. 
Thus the instability is a pure relativistic effect ( the same is verified for circular orbits around the superposition of black 
holes and axially symmetric structures in Weyl coordinates, see \citet{b42}).
\subsection{Second Model} \label{sub_sec_miy2}

Another pair of potential and density functions presented by Miyamoto and Nagai as a generalization of 
Toomre's model 2 is described by
\begin{subequations} 
\begin{align}
\Phi(R,z) &= -\frac{GM}{\sqrt{R^2+(a+\zeta)^2}} \left[ 1+\frac{a(a+\zeta)}{R^2+(a+\zeta)^2} \right] \mbox{,} \\
\rho_N(R,z) &=\frac{3b^2M}{4\pi\zeta^3\left[ R^2+(a+\zeta)^2 \right]^{7/2}} \left[ R^2(a^3+\zeta^3) 
+(\zeta+a)^3(\zeta^2+4a\zeta+a^2) \right] \mbox{,}
\end{align}
\end{subequations}
where $\zeta=\sqrt{z^2+b^2}$. Using $f$ defined by Eq.\ (\ref{eq_f}) in 
Eq.\ (\ref{eq_Ttt})--(\ref{eq_Tfifi}), we find that the components $T^R_z=T^z_R$ are 
nonzero. The physical variables of the matter distribution are obtained by 
solving the eigenvalue problem for $T^a_b$
\begin{equation}
T^a_b\Xi^b=\lambda\Xi^a \mbox{.}
\end{equation}
We find that $T^{ab}$ can be put in the form
\begin{equation} \label{eq_emt_miy2}
T^{ab}=\sigma e_{(0)}^ae_{(0)}^b+p_{+} e_{(1)}^ae_{(1)}^b+p_{-}
e_{(2)}^ae_{(2)}^b+p_{\varphi}e_{(3)}^ae_{(3)}^b \mbox{,}
\end{equation}
where
\begin{align}
\sigma &=\frac{T^t_t}{c^2}, \qquad e_{(0)}^a=\left(\frac{1+f}{1-f},0,0,0\right) \mbox{,} \notag \\
p_{\pm} &=-\frac{T^R_R+T^z_z}{2} \mp \frac{1}{2}\sqrt{(T^R_R-T^z_z)^2+4(T^R_z)^2}
\mbox{,} \label{eq_p+-} \\
e_{(1)}^a &=(0,e_{(1)}^R,e_{(1)}^z,0), \qquad e_{(2)}^a=(0,e_{(2)}^R,e_{(2)}^z,0)  \mbox{,} \notag \\
p_{\varphi} &= -T^{\varphi}_{\varphi}, \qquad e_{(3)}^a=\left(0,0,0,\frac{1}{R(1+f)^2} \right) \mbox{,}
\end{align}
and 
\begin{align}
e_{(1)}^R &=-\frac{T^R_z}{(1+f)^2\sqrt{(T^R_z)^2+(T^R_R+p_{+})^2}} \mbox{,} \qquad
e_{(1)}^z = \frac{T^R_R+p_{+}}{(1+f)^2\sqrt{(T^R_z)^2+(T^R_R+p_{+})^2}} \mbox{,} \notag \\
e_{(2)}^R &= -\frac{T^R_z}{(1+f)^2\sqrt{(T^R_z)^2+(T^R_R+p_{-})^2}} \mbox{,} \qquad
e_{(2)}^z = \frac{T^R_R+p_{-}}{(1+f)^2\sqrt{(T^R_z)^2+(T^R_R+p_{-})^2}} \mbox{.}
\end{align}
The effective Newtonian density reads $\rho=\sigma+p_{+}/c^2+p_{-}/c^2+p_{\varphi}/c^2=
\sigma-T^R_R/c^2-T^z_z/c^2+p_{\varphi}/c^2$. The explicit expressions for $\bar{\sigma}$, 
$\bar{\rho}$ and $\bar{p}_{\varphi}$ are
\begin{align}
\bar{\sigma} &=\frac{3\tilde{b}^2 \chi^4 \left[ \tilde{R}^2(\tilde{a}^3+\xi ^3)+
(\tilde{a}+\xi)^3(\tilde{a}^2+\xi
 ^2+4\tilde{a} \xi) \right]}{4 \pi \xi ^3 \left[
 \chi^{3/2} +\tilde{R}^2+ (\tilde{a}+\xi)(2\tilde{a}+\xi )\right]^5} \mbox{,} \label{eq_sigma_miy2}\\
\bar{\rho} &= \frac{3\tilde{b}^2 \chi^{11/2} \left[ \tilde{R}^2(\tilde{a}^3+\xi^3)+(\tilde{a}+\xi)^3(\tilde{a}^2+\xi^2+
4\tilde{a} \xi ) \right]}{4\pi \xi^3 \left[ \chi^{3/2} +\tilde{R}^2+ (\tilde{a}+\xi)(2\tilde{a}+\xi ) \right]^5
\left[ \chi^{3/2} -\tilde{R}^2-(\tilde{a}+\xi)(2\tilde{a}+\xi ) \right]} \mbox{,} \label{eq_rho_miy2} \\
\bar{p}_{\varphi} &= \frac{\chi^4}{8 \pi \xi ^3
\left[ \chi^{3/2} +\tilde{R}^2+ (\tilde{a}+\xi)(2\tilde{a}+\xi ) \right]^5
\left[ \chi^{3/2} -\tilde{R}^2-(\tilde{a}+\xi)(2\tilde{a}+\xi ) \right]}
 \left\{ \tilde{R}^4 \left[ \xi ^3(\tilde{a}^2+2\tilde{b}^2)+3\tilde{a}^3\tilde{b}^2\right] \right. \notag \\
& \left. +\tilde{R}^2(\tilde{a}+\xi)^2\left[ 9\tilde{a}^3\tilde{b}^2
+12\tilde{a}\tilde{b}^2\xi^2+4\tilde{b}^2\xi^3+2\tilde{a}^2(7\tilde{b}^2+\tilde{z}^2)\xi \right]
+(a+\xi)^4 \left[ 6\tilde{a}^3\tilde{b}^2+12\tilde{a}\tilde{b}^2 \xi^2+2\tilde{b}^2\xi^3+
\tilde{a}^2(19\tilde{b}^2+\tilde{z}^2)\xi \right] \right\} \mbox{,} \label{eq_pfi_miy2}
\end{align}
where $\xi=\sqrt{\tilde{z}^2+\tilde{b}^2}$, $\chi=\tilde{R}^2+(\tilde{a}+\xi)^2$ and the 
dimensionless variables were previously defined. The components $T^R_R$, $T^R_z$ and 
$T^z_z$ read
\begin{align}
T^R_R &=-\frac{Mc^2}{r_s^3}\frac{\chi^4}{8\pi \xi^3 \left[ \chi^{3/2} +\tilde{r}^2+ (\tilde{a}+\xi)(2\tilde{a}+\xi ) \right]^5
\left[ \chi^{3/2} -\tilde{r}^2-(\tilde{a}+\xi)(2\tilde{a}+\xi ) \right]} \left\{ \tilde{r}^4 \left[ \xi^3(\tilde{a}^2+2\tilde{b}^2)
+3\tilde{a}^3\tilde{b}^2 \right] \right. \notag \\
& \left. +\tilde{r}^2(\tilde{a}+\xi)^2 \left[ 9\tilde{a}^3\tilde{b}^2+12\tilde{a}\tilde{b}^2\xi^2+4\tilde{b}^2\xi^3
+2\tilde{a}^2(\tilde{b}^2-5\tilde{z}^2)\xi \right]+(\tilde{a}+\xi)^4\left[ 6\tilde{a}^3\tilde{b}^2
+12\tilde{a}\tilde{b}^2 \xi^2+2\tilde{b}^2\xi^3+\tilde{a}^2(19\tilde{b}^2+\tilde{z}^2)\xi \right] \right\} \mbox{,} \\
T^R_z &=T^z_R = -\frac{Mc^2}{r_s^3}\frac{3\tilde{a}^2\tilde{r}\tilde{z}(\tilde{a}+\xi) \chi^4\left[ \tilde{r}^2-(a+\xi)^2\right]}{4\pi \xi
\left[ \chi^{3/2} +\tilde{r}^2+ (\tilde{a}+\xi)(2\tilde{a}+\xi ) \right]^5
\left[ \chi^{3/2} -\tilde{r}^2-(\tilde{a}+\xi)(2\tilde{a}+\xi ) \right]} \mbox{,} \\
T^z_z &=\frac{Mc^2}{r_s^3}\frac{\chi^4}{4\pi \xi^2 \left[ \chi^{3/2} +\tilde{r}^2+ (\tilde{a}+\xi)(2\tilde{a}+\xi ) \right]^5
\left[ \chi^{3/2} -\tilde{r}^2-(\tilde{a}+\xi)(2\tilde{a}+\xi ) \right]} \times \notag \\
& \left\{ \tilde{r}^4\xi^2(\tilde{a}^2-\tilde{b}^2)
-2\tilde{r}^2\xi(\tilde{a}+\xi)^2
\left[ 3\tilde{a}\tilde{b}^2+\xi(2\tilde{a}^2+\tilde{b}^2) \right] +(\tilde{a}+\xi)^4 \left[ \tilde{a}^2
(\tilde{z}^2-8\tilde{b}^2)-\tilde{b}^2\xi (6\tilde{a}+\xi) \right] \right\} \mbox{,} 
\end{align}
and $p_{\pm}$ is further adimensionalized as $p_{\pm}=\frac{Mc^2}{r_s^3}\bar{p}_{\pm}$.  

\begin{figure}
\centering
\includegraphics[scale=0.5]{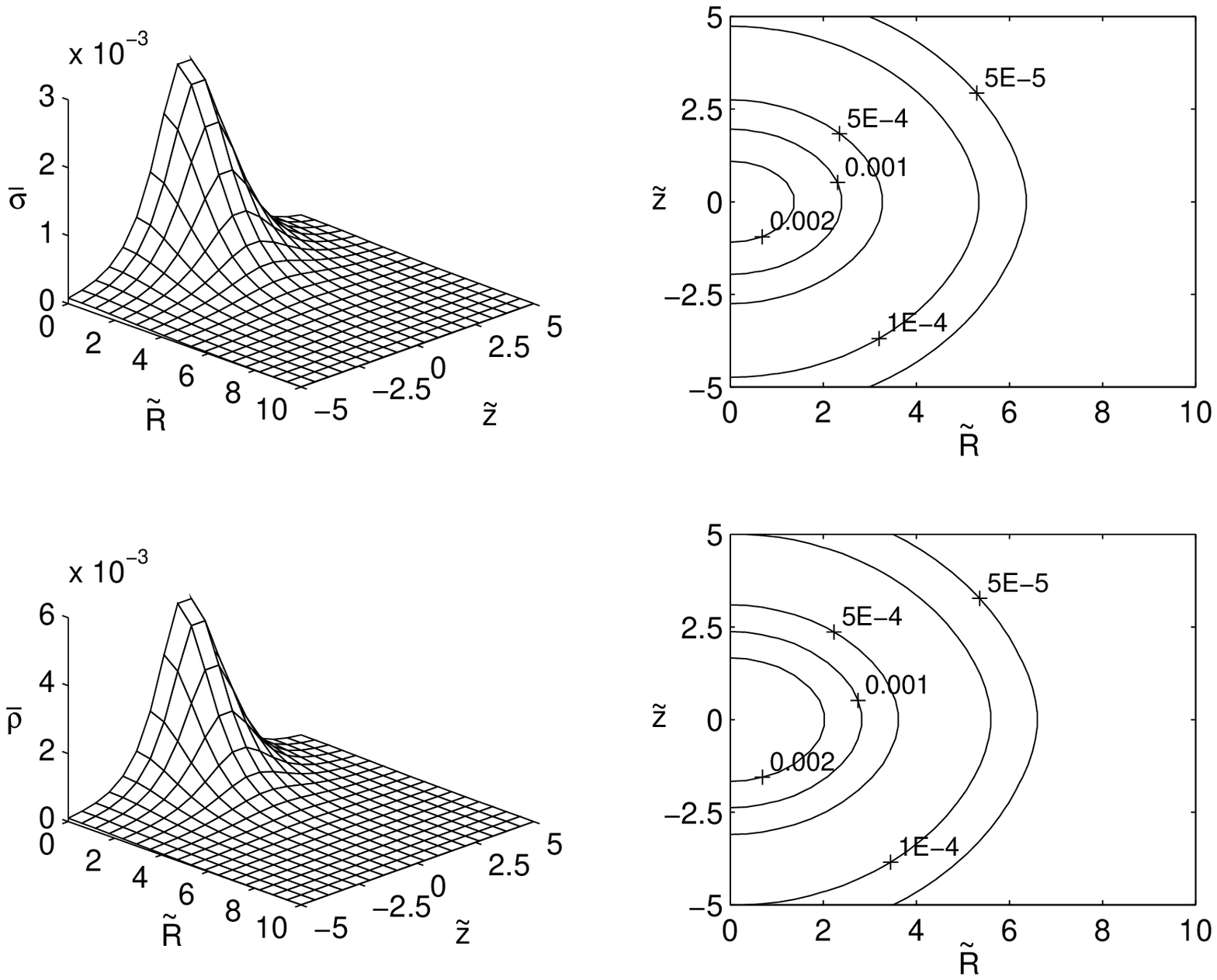}
\caption{Surface plot and level curves of the energy density $\bar{\sigma}$ Eq.\ (\ref{eq_sigma_miy2}) and
``effective Newtonian density'' $\bar{\rho}$ Eq.\ (\ref{eq_rho_miy2}) with parameters $\tilde{a}=1$ and $\tilde{b}=2$.} \label{fig6}
\end{figure}

\begin{figure}
\centering
\includegraphics[scale=0.5]{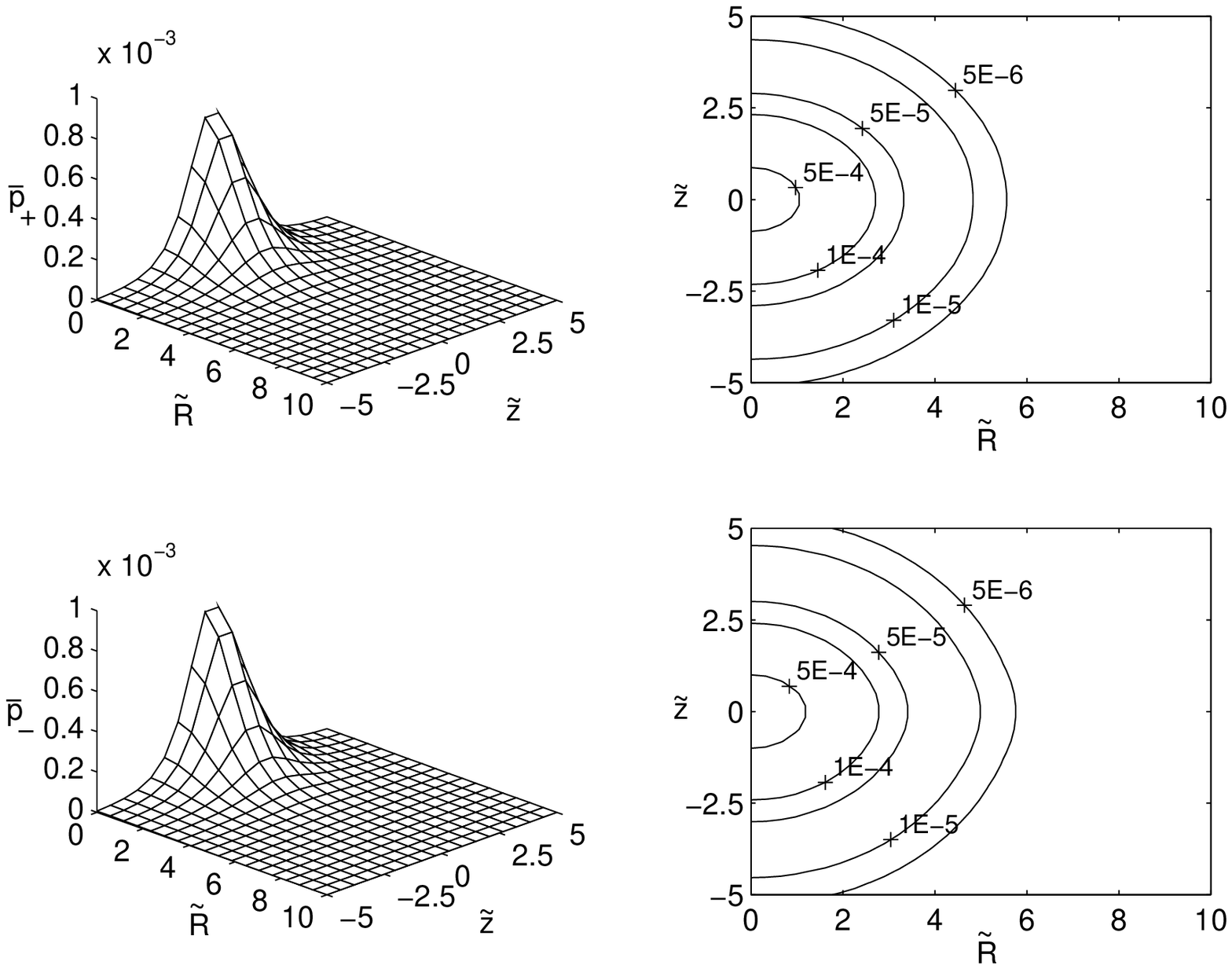}
\caption{Surface plot and level curves of the pressures $\bar{p}_{\pm}$ Eq.\ (\ref{eq_p+-})
with parameters $\tilde{a}=1$ and $\tilde{b}=2$.} \label{fig7}
\end{figure}

\begin{figure}
\centering
\includegraphics[scale=0.5]{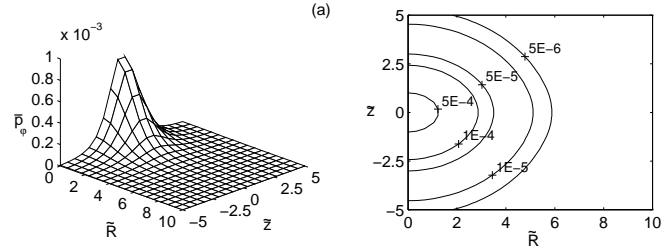}
\caption{Surface plot and level curves of the azimuthal pressure $\bar{p}_{\varphi}$ Eq.\ (\ref{eq_pfi_miy2})
with parameters $\tilde{a}=1$ and $\tilde{b}=2$.} \label{fig8}
\end{figure}

\begin{figure}
\centering
\includegraphics[scale=0.68]{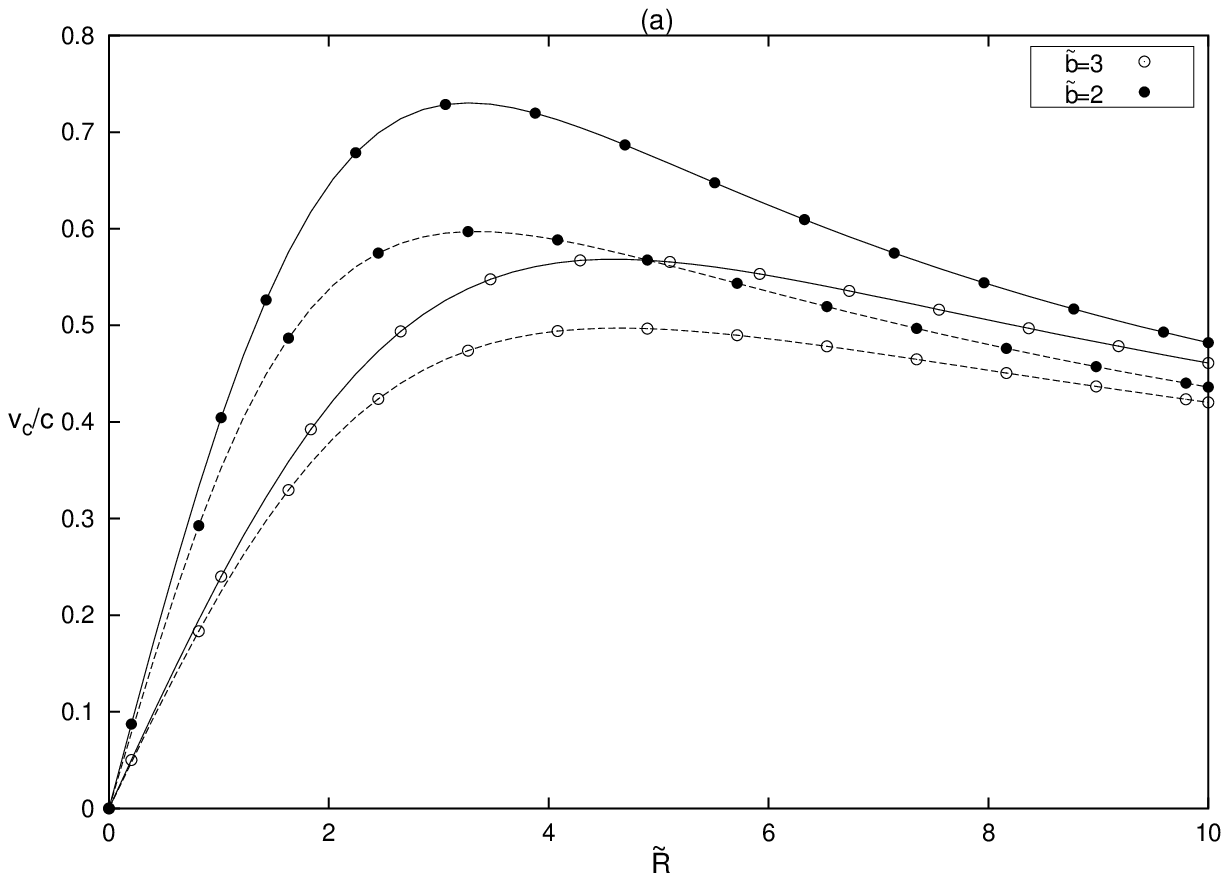}\\
\vspace{0.2cm}
\includegraphics[scale=0.68]{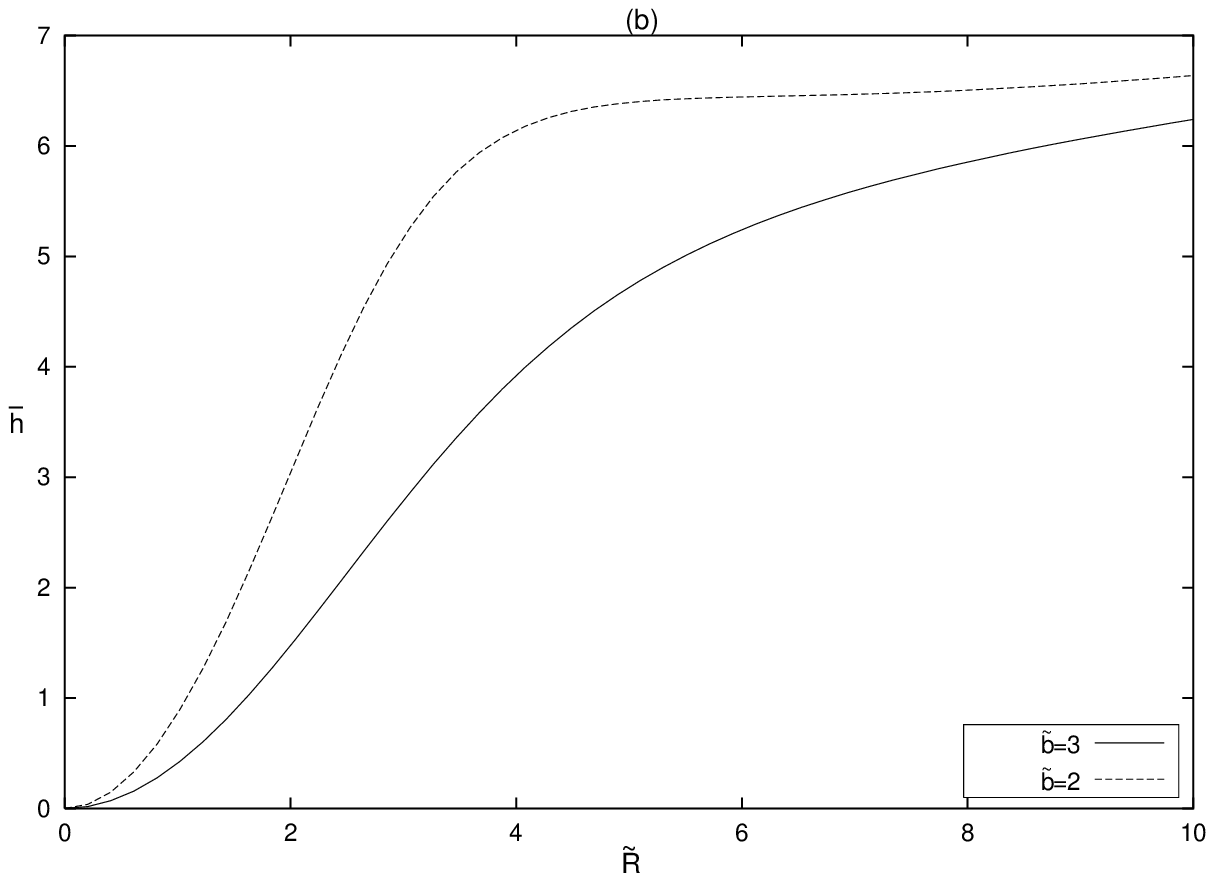}
\caption{(a) Relativistic rotation profile Eq.\ (\ref{eq_vc_miy2}) and Newtonian rotation profile
(dashed curves) with parameters $\tilde{a}=1$ and $\tilde{b}=3$, $\tilde{b}=2$. (b)
Specific angular momentum Eq.\ (\ref{eq_h_miy2}) for the same parameters.} \label{fig9}
\end{figure}

The condition $(\tilde{a}+\tilde{b})^2>(2\tilde{a}+\tilde{b})$ ensures positive densities and
non-singular behaviour for the densities and pressures. The first order expansion of 
Eq.\ (\ref{eq_rho_miy2}) reads
\begin{equation}
\bar{\rho}  \approx \frac{3\tilde{b}^2\left[ \tilde{R}^2(\tilde{a}^3+\xi^3)+
(\tilde{a}+\xi)^3(\tilde{a}^2+\xi^2+4\tilde{a}\xi) \right]}{4\pi \xi^3 [\tilde{R}^2+(\tilde{a}+\xi)^2]^{7/2}} 
 \left\{ 1-\frac{4[\tilde{R}^2+(\tilde{a}+\xi)(2\tilde{a}+\xi)]}{[\tilde{R}^2+(\tilde{a}+\xi)^2]^{3/2}} \right\} \mbox{.}
\end{equation}
The surfaces and level curves of the densities Eq.\ (\ref{eq_sigma_miy2})--(\ref{eq_rho_miy2}) and 
pressures Eq.\ (\ref{eq_p+-}) are shown in Fig.\ \ref{fig6} and Fig.\ \ref{fig7}, respectively, for
parameters $\tilde{a}=1$, $\tilde{b}=2$. Fig.\ \ref{fig8}(a) displays the surface and level curves 
for the azimuthal pressure. For these parameters we have $\tilde{p}_{\varphi}/\tilde{\sigma} < 0.3$,
$\tilde{p}_{+}/\tilde{\sigma}< 0.25$ and $\tilde{p}_{-}/\tilde{\sigma} < 0.3$.

The tangential circular velocity Eq.\ (\ref{eq_vcf}) and specific angular momentum Eq.\ (\ref{eq_hf}) 
are given by
\begin{align}
& \left( \frac{v_c}{c} \right)^2 =\frac{2\tilde{R}^2\chi^{3/2}\left[ \tilde{R}^2+\eta
(4\tilde{a}+\tilde{b})\right]}{\left[ \chi^{3/2}-\tilde{R}^2- \eta(2\tilde{a}+\tilde{b})\right]
\left[ \chi^{5/2}-\tilde{R}^4-5\tilde{a}\eta\tilde{R}^2+\eta^3(2\tilde{a}+\tilde{b})\right]}\mbox{,} \label{eq_vc_miy2} \\
& \bar{h}=\frac{\sqrt{2}\tilde{R}^2\left[ \chi^{3/2}+\tilde{R}^2+\eta(2\tilde{a}+\tilde{b})\right]^2
\sqrt{\tilde{R}^2+\eta(4\tilde{a}+\tilde{b})}}{\chi^{9/4}} 
 \left\{\chi^4+ \left[ \tilde{R}^2+\eta(2\tilde{a}+\tilde{b}) \right]\left[ \tilde{R}^4+5\tilde{a}\eta\tilde{R}^2
-\eta^3(2\tilde{a}+\tilde{b})\right] \right. \notag \\
& \left. -4 \tilde{R}^2\chi^{3/2}
\left[ \tilde{R}^2+\eta(4\tilde{a}+\tilde{b})\right] \right\}^{-1/2} \mbox{,} \label{eq_h_miy2}
\end{align}
where $h=cr_s\bar{h}$, $\eta=\tilde{a}+\tilde{b}$ and $\chi=\tilde{R}^2+\eta^2$. The first
order expansion of Eq.\ (\ref{eq_vc_miy2}) reads
\begin{equation} \label{eq_vc_miy2c}
\frac{v_c}{c} =\frac{\sqrt{2}\tilde{R}\left[ \tilde{R}^2+(\tilde{a}+\tilde{b})(4\tilde{a}+\tilde{b})\right]^{1/2}}
{\left[ \tilde{R}^2+(\tilde{a}+\tilde{b})^2 \right]^{5/4}} 
\left\{
1+\frac{\tilde{R}^2\left[ \tilde{R}^2+(\tilde{a}+\tilde{b})(4\tilde{a}+\tilde{b})\right]}
{\left[ \tilde{R}^2+(\tilde{a}+\tilde{b})^2 \right]^{5/2}} \right\} \mbox{.}
\end{equation}
In Fig.\ \ref{fig9}(a)--(b) we display the rotation profiles and curves of specific angular momentum
for parameters $\tilde{a}=1$, $\tilde{b}=3$ and $\tilde{b}=2$. As in the previous case, relativistic effects 
increase the tangential velocity. We also note that the orbits for $\tilde{b}=3$ and $\tilde{b}=2$ are 
all stable. 
\section{A General Relativistic Satoh Model} \label{sec_satoh}

\citet{b7} derived other pairs of three-dimensional potential and density functions by inflating the 
Plummer models of order $n$. In the limit when $n \rightarrow \infty$ the pair takes the simple form
\begin{subequations}
\begin{align}
\Phi(R,z) &=-\frac{Gm}{\sqrt{R^2+z^2+a(a+2\sqrt{z^2+b^2})}} \mbox{,} \label{eq_phin_satoh}\\
\rho_N(R,z) &=\frac{ab^2M}{4\pi} 
 \frac{\left[ R^2+z^2+(a+2\sqrt{z^2+b^2})(a+3\sqrt{z^2+b^2}) \right]}
{(z^2+b^2)^{3/2}\left[ R^2+z^2+a(a+2\sqrt{z^2+b^2}) \right]^{5/2}} \mbox{.}
\end{align}
\end{subequations}
Using the function $f(R,z)$ with $\Phi$ given by Eq.\ (\ref{eq_phin_satoh}), we get the 
following expressions for the nonzero components of the energy-momentum tensor 
Eq.\ (\ref{eq_Ttt})--(\ref{eq_Tfifi})
\begin{subequations}
\begin{align}
\bar{\sigma} &=\frac{\tilde{a}\tilde{b}^2 \left[ \tilde{R}^2+\tilde{z}^2+(\tilde{a}+2\xi)(\tilde{a}+3\xi) \right]}
{8\pi\xi^3\left[ 1+\sqrt{\tilde{R}^2+\tilde{z}^2+\tilde{a}(\tilde{a}+2\xi)} \right]^5} 
\mbox{,} \label{eq_sigma_satoh}\\
\bar{p}_R &= \bar{p}_{\varphi}=\frac{\tilde{a}\tilde{b}^2\left[ \tilde{R}^2+\tilde{z}^2+(\tilde{a}+2\xi)^2 \right]}
{8\pi\xi^3\left[ 1+\sqrt{\tilde{R}^2+\tilde{z}^2+\tilde{a}(\tilde{a}+2\xi)} \right]^5 \left[
-1+\sqrt{\tilde{R}^2+\tilde{z}^2+\tilde{a}(\tilde{a}+2\xi)} \right]} \mbox{,} \label{eq_pr_satoh}\\
\bar{p}_z &=\frac{\tilde{a}\tilde{b}^2(\tilde{a}+2\xi)}{4\pi\xi^2
\left[ 1+\sqrt{\tilde{R}^2+\tilde{z}^2+\tilde{a}(\tilde{a}+2\xi)} \right]^5 \left[
-1+\sqrt{\tilde{R}^2+\tilde{z}^2+\tilde{a}(\tilde{a}+2\xi)} \right]} \mbox{,} \label{eq_pz_satoh}\\
\bar{\rho} &=\frac{\tilde{a}\tilde{b}^2\sqrt{\tilde{R}^2+\tilde{z}^2+\tilde{a}(\tilde{a}+2\xi)}
\left[ \tilde{R}^2+\tilde{z}^2+(\tilde{a}+2\xi)(\tilde{a}+3\xi) \right]}{4\pi\xi^3
\left[ 1+\sqrt{\tilde{R}^2+\tilde{z}^2+\tilde{a}(\tilde{a}+2\xi)} \right]^5 \left[
-1+\sqrt{\tilde{R}^2+\tilde{z}^2+\tilde{a}(\tilde{a}+2\xi)} \right]} \mbox{,} \label{eq_rho_satoh}
\end{align}
\end{subequations}
where $\xi=\sqrt{\tilde{z}^2+\tilde{b}^2}$ and the dimensionless variables were previously 
defined. As $f$ has the form of Eq.\ (\ref{eq_f_best}), the energy-momentum tensor is diagonal 
and the radial and azimuthal stresses are equal. We have $\bar{\sigma}>0$ and $\bar{\rho}>0$ 
everywhere and pressures without
 singularities provided $\tilde{a}(\tilde{a}+2\tilde{b})>1$. 
In the weak field limit, Eq.\ (\ref{eq_rho_satoh}) may be expanded as
\begin{equation}
\bar{\rho}  \approx \frac{\tilde{a}\tilde{b}^2}{4\pi}\frac{\left[ \tilde{R}^2+\tilde{z}^2+
(\tilde{a}+2\xi)(\tilde{a}+3\xi) \right]}
{\xi^3\left[ \tilde{R}^2+\tilde{z}^2+\tilde{a}(\tilde{a}+2\xi) \right]^{5/2}} 
\left[ 1-\frac{4}{\sqrt{\tilde{R}^2+\tilde{z}^2+\tilde{a}(\tilde{a}+2\xi)}} \right] \mbox{,}
\end{equation}
which introduces again a negative first order relativistic correction to the Newtonian mass density.
\begin{figure}
\centering
\includegraphics[scale=0.5]{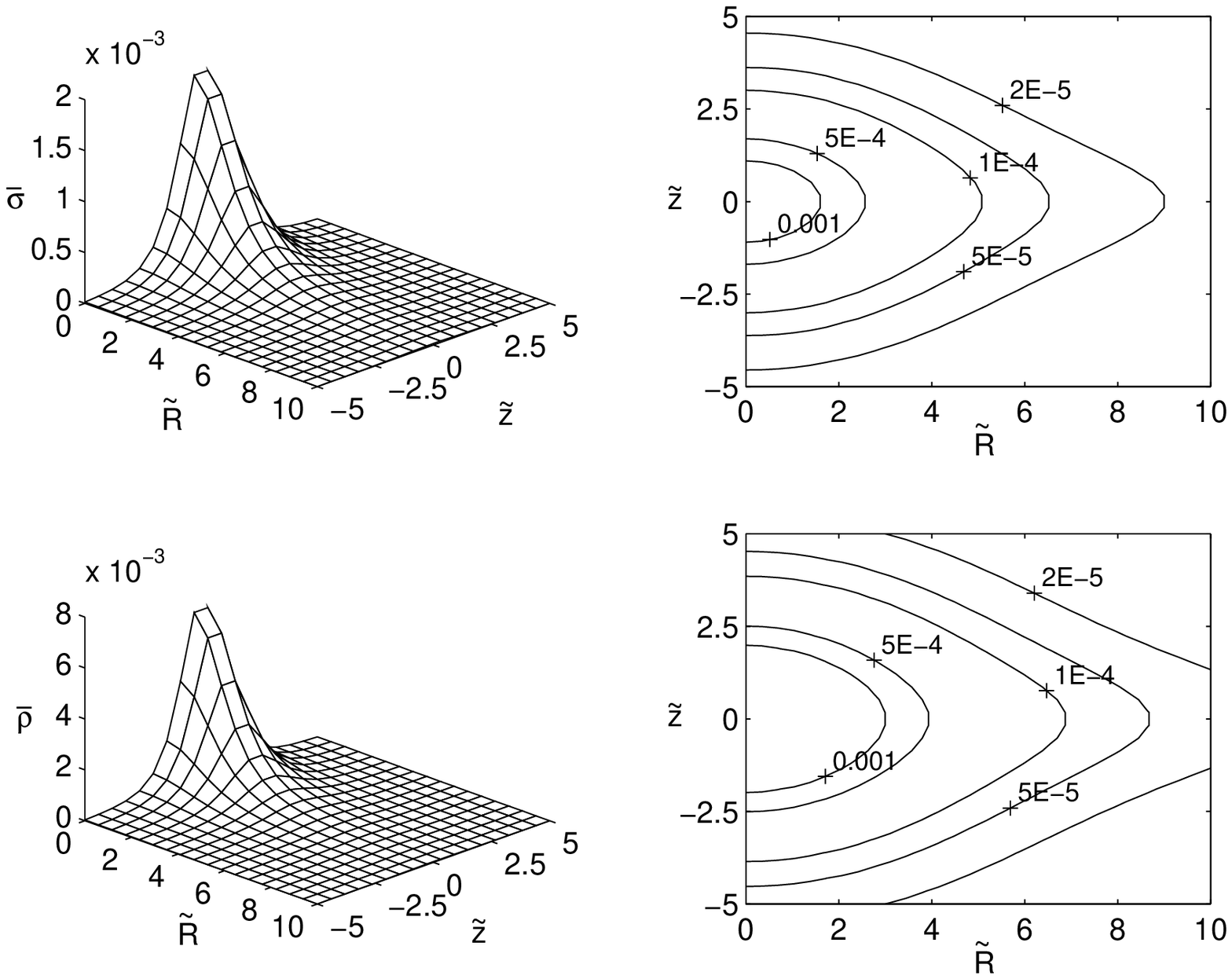}
\caption{Surface plot and level curves of the energy density $\bar{\sigma}$ Eq.\ (\ref{eq_sigma_satoh}) and ``effective
Newtonian density'' $\bar{\rho}$ Eq.\ (\ref{eq_rho_satoh}) with parameters $\tilde{a}=1$ and $\tilde{b}=2$.} \label{fig10}
\end{figure}

\begin{figure}
\centering
\includegraphics[scale=0.5]{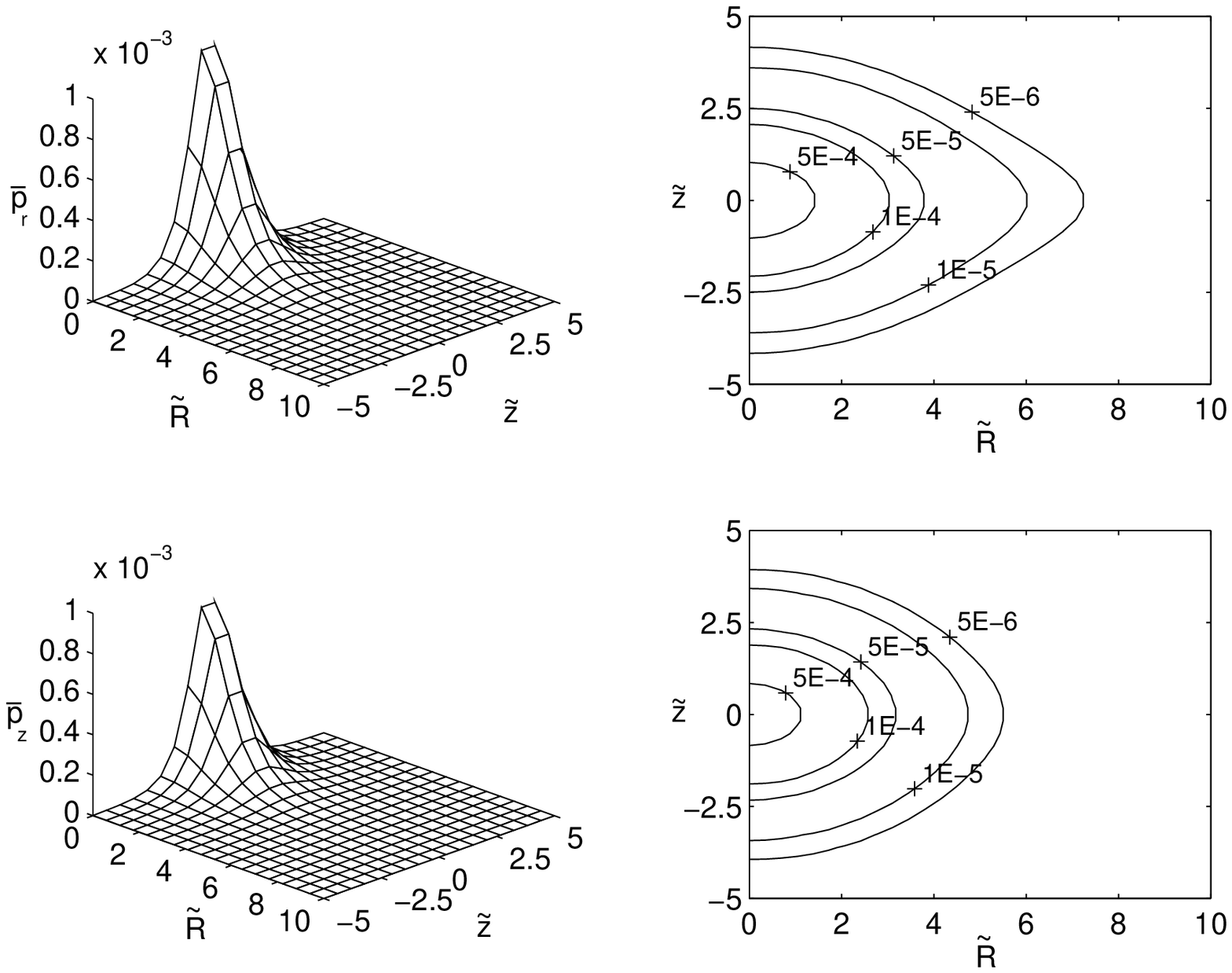}
\caption{Surface plot and level curves of the radial pressure $\bar{p}_{R}$ Eq.\ (\ref{eq_pr_satoh}) and vertical
pressure $\bar{p}_{z}$ Eq.\ (\ref{eq_pz_satoh}) with parameters $\tilde{a}=1$ and $\tilde{b}=2$.} \label{fig11}
\end{figure}
Expressions for the tangential velocity and angular momentum follow directly from 
Eq.\ (\ref{eq_vcf})--(\ref{eq_hf})
\begin{align}
& \left( \frac{v_c}{c} \right)^2 =\frac{2\tilde{R}^2\sqrt{\tilde{R}^2+\eta}}{\left[
-\tilde{R}^2+\eta+( \tilde{R}^2+\eta)^{3/2} \right]
\left( -1+\sqrt{\tilde{R}^2+\eta} \right)} \mbox{,} \label{eq_vc_satoh}\\
& \bar{h} =\frac{\sqrt{2}\tilde{R}^2 \left(1+\sqrt{\tilde{R}^2+\eta}\right)^2}{(\tilde{R}^2+\eta)^{3/4}
\sqrt{ (\tilde{R}^2+\eta)^2+\tilde{R}^2-\eta-4\tilde{R}^2\sqrt{\tilde{R}^2+\eta}}} \mbox{,} \label{eq_h_satoh}
\end{align}
where $h=cr_s\bar{h}$ and $\eta=\tilde{a}(\tilde{a}+2\tilde{b})$. The first order expansion of Eq.\ (\ref{eq_vc_satoh}) is given by
\begin{equation}
\frac{v_c}{c}=\frac{\tilde{R}\sqrt{2}}{\left[ \tilde{R}^2+\tilde{a}(\tilde{a}+2\tilde{b}) \right]^{3/4}} \left\{
1+\frac{\tilde{R}^2}{\left[ \tilde{R}^2+\tilde{a}(\tilde{a}+2\tilde{b}) \right]^{3/2}} \right\} \mbox{.}
\end{equation}

\begin{figure}
\centering
\includegraphics[scale=0.68]{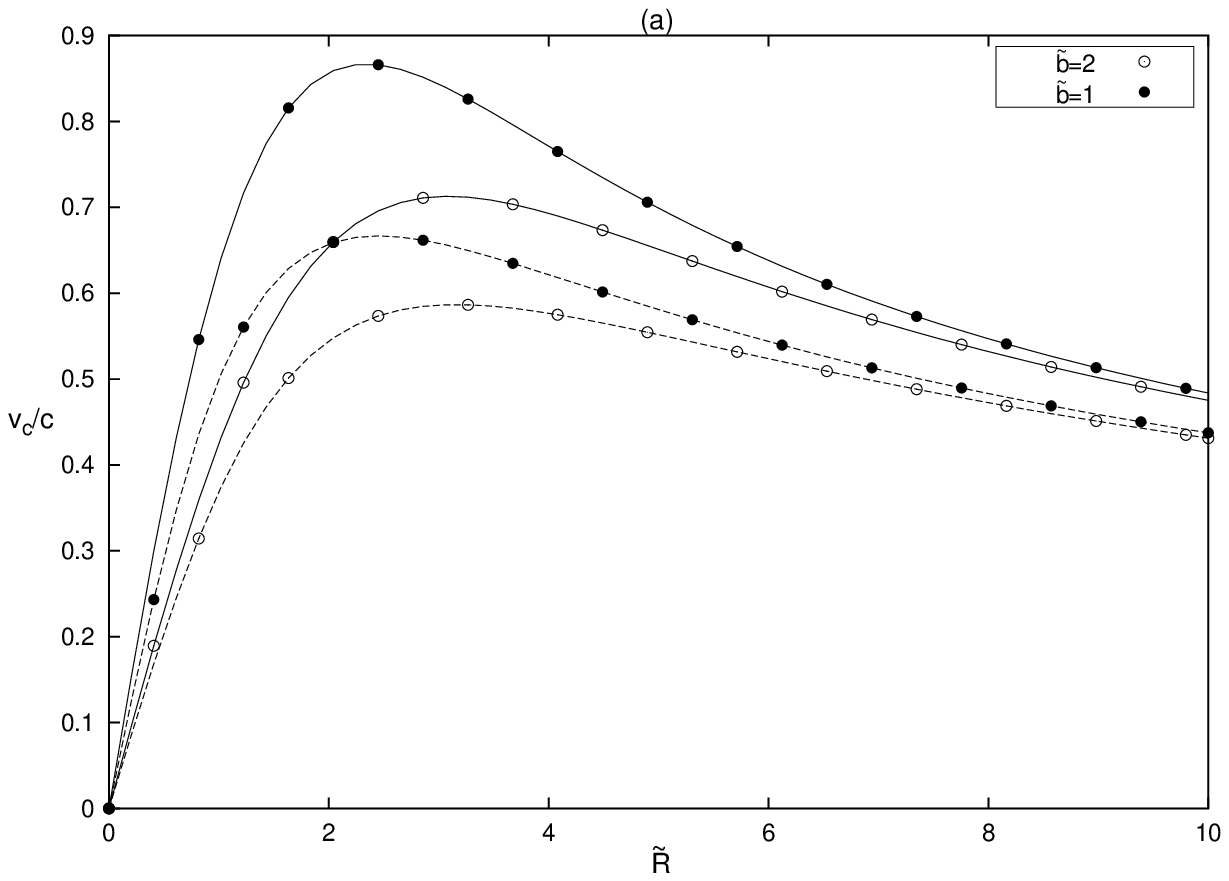}\\
\vspace{0.2cm}
\includegraphics[scale=0.68]{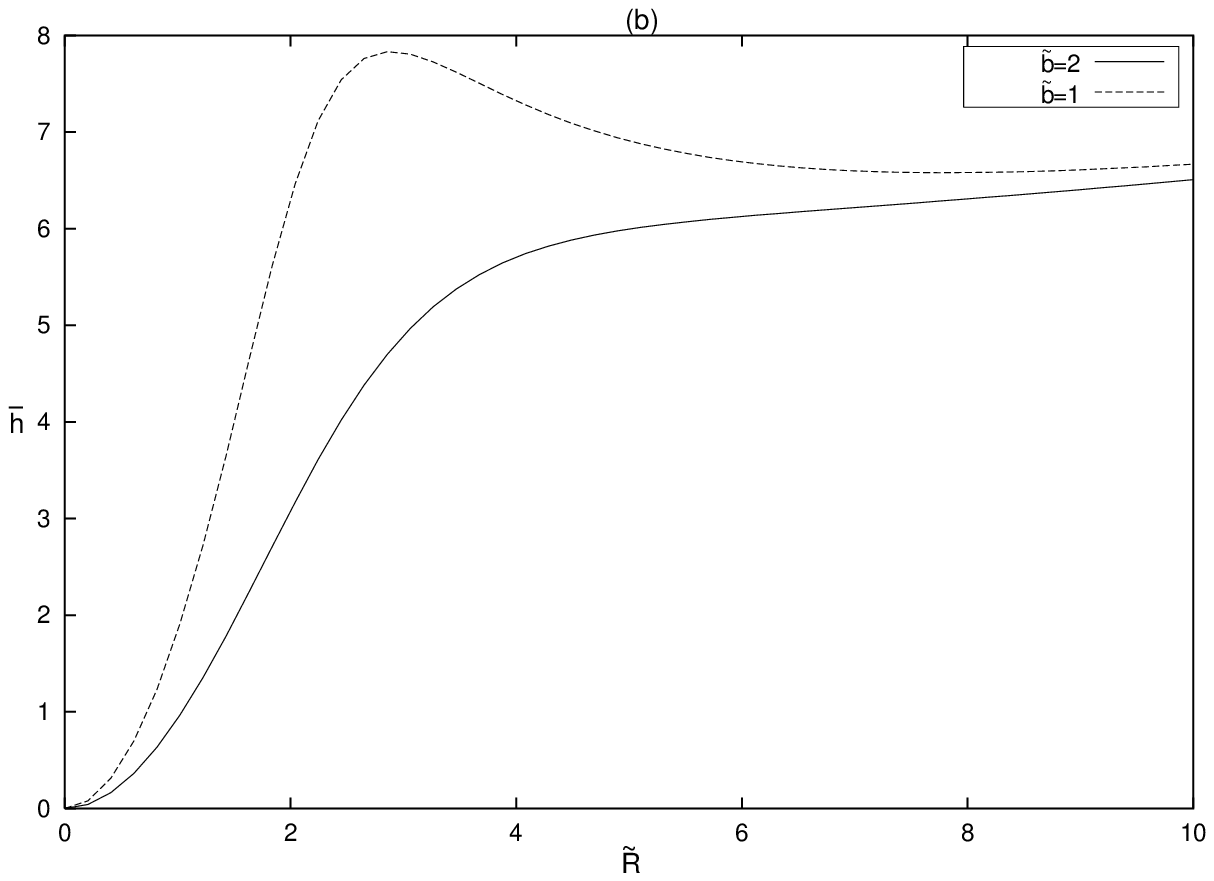}
\caption{(a) Relativistic rotation profile Eq.\ (\ref{eq_vc_satoh}) and Newtonian rotation profile
(dashed curves) with parameters $\tilde{a}=1$ and $\tilde{b}=2$, $\tilde{b}=1$. (b)
Specific angular momentum Eq.\ (\ref{eq_h_satoh}) for the same parameters.} \label{fig12}
\end{figure}

The surfaces and level curves of the densities and pressures
Eq.\ (\ref{eq_sigma_satoh})--(\ref{eq_rho_satoh}) for parameters $\tilde{a}=1$, $\tilde{b}=2$ are plotted in
Fig.\ \ref{fig10} and Fig.\ \ref{fig11}. We have $\bar{p}_R/\bar{\sigma} <0.6$ and $\bar{p}_z/\bar{\sigma} <0.5$,
so all energy conditions are satisfied. Rotation profiles and curves of specific angular momentum are
displayed in Fig.\ \ref{fig12} for parameters $\tilde{a}=1$, $\tilde{b}=2$ and $\tilde{b}=1$. We note that
circular orbits are stable for $\tilde{b}=2$.
\section{Geodesic Orbits} \label{sec_geo}

An interesting application of the General Relativistic galactic models is the
study of geodesic orbits outside the galactic plane. As an example we 
calculate numerically two orbits for the first Miyamoto-Nagai model 
(Sec.\ \ref{sub_sec_miy1}) and compare them with the Newtonian case. 

We solve numerically the geodesic equations of motion
\begin{equation} \label{eq_geo}
\ddot{x}^{\mu}+\Gamma^{\mu}_{\alpha \beta}\dot{x}^{\alpha}\dot{x}^{\beta}=0 \mbox{,}
\end{equation}
for metric Eq.\ (\ref{eq_metr_iso}), where $\Gamma^{\mu}_{\alpha \beta}$ are 
the Christoffel symbols and the dot denote differentiation with respect to the
proper time. With the orthonormal tetrad defined as
\begin{align}
e_{(t)}^i &=\frac{1+f}{1-f}(1,0,0,0), \quad e_{(R)}^i =\frac{1}{(1+f)^2}(0,1,0,0) \mbox{,} \notag \\
e_{(z)}^i &=\frac{1}{(1+f)^2}(0,0,1,0), \quad e_{(\varphi)}^i =\frac{1}{R(1+f)^2}(0,0,0,1) \mbox{,}
\end{align}
the tetrad components of the four-velocity $v^i$ read
\begin{equation}
v^i=\gamma(1,v\sin \psi \cos \theta,v\sin \psi \sin \theta,v\cos \psi) \mbox{,}
\end{equation}
with $\gamma=1/\sqrt{1-v^2}$ (we have set $c=1$). The specific energy and angular momentum of the test
particle are
\begin{align}
\mathcal{E} &=\gamma \frac{1-f}{1+f}\mbox{,} \\
h &=r(1+f)^2\gamma v \cos \psi \mbox{.}
\end{align}
As initial conditions we take a position at radius $R_0$ on the disk's plane and components of the
four-velocity $v_0^i=\gamma(1,0 ,v_0\sin \psi,v_0\cos \psi)$, where $v_0$ is equal to
the tangential velocity of circular orbits at radius $R_0$. Similar initial conditions for the Newtonian 
case are chosen with the same angular momentum and energy $\mathcal{E}-1$. Both the geodesic 
and Newtonian equations are solved with the fourth-fifth Runge-Kutta method with adaptive size step.
\begin{figure}
\centering
\includegraphics[scale=0.5]{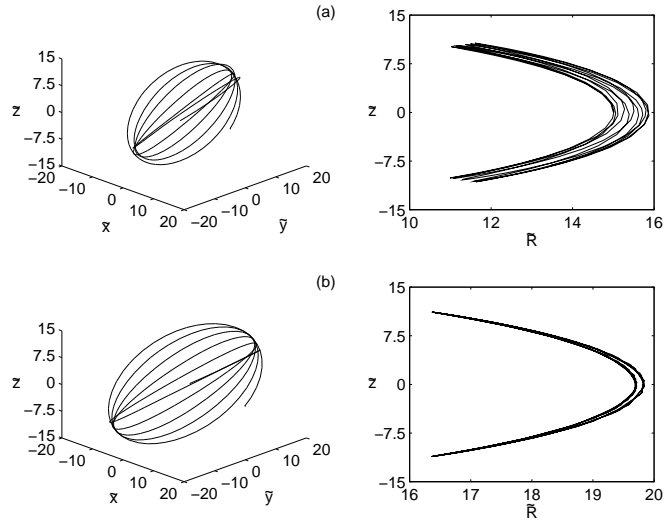}
\caption{(a) Geodesic orbits for the first Miyamoto-Nagai model with
parameters $\tilde{a}=1$, $\tilde{b}=1$. Initial conditions: $\tilde{R}_0=15$,
$\psi=45^o$ with energy $\mathcal{E}\approx 0.901$  and angular momentum $\bar{h} \approx 5.03$. (b) Newtonian orbit
with energy $\mathcal{E}-1$ and same angular momentum. Initial conditions: $\tilde{R}_0 
\approx 19.84$ and $\psi \approx 36.37^o$.} \label{fig13}
\end{figure}

\begin{figure}
\centering
\includegraphics[scale=0.5]{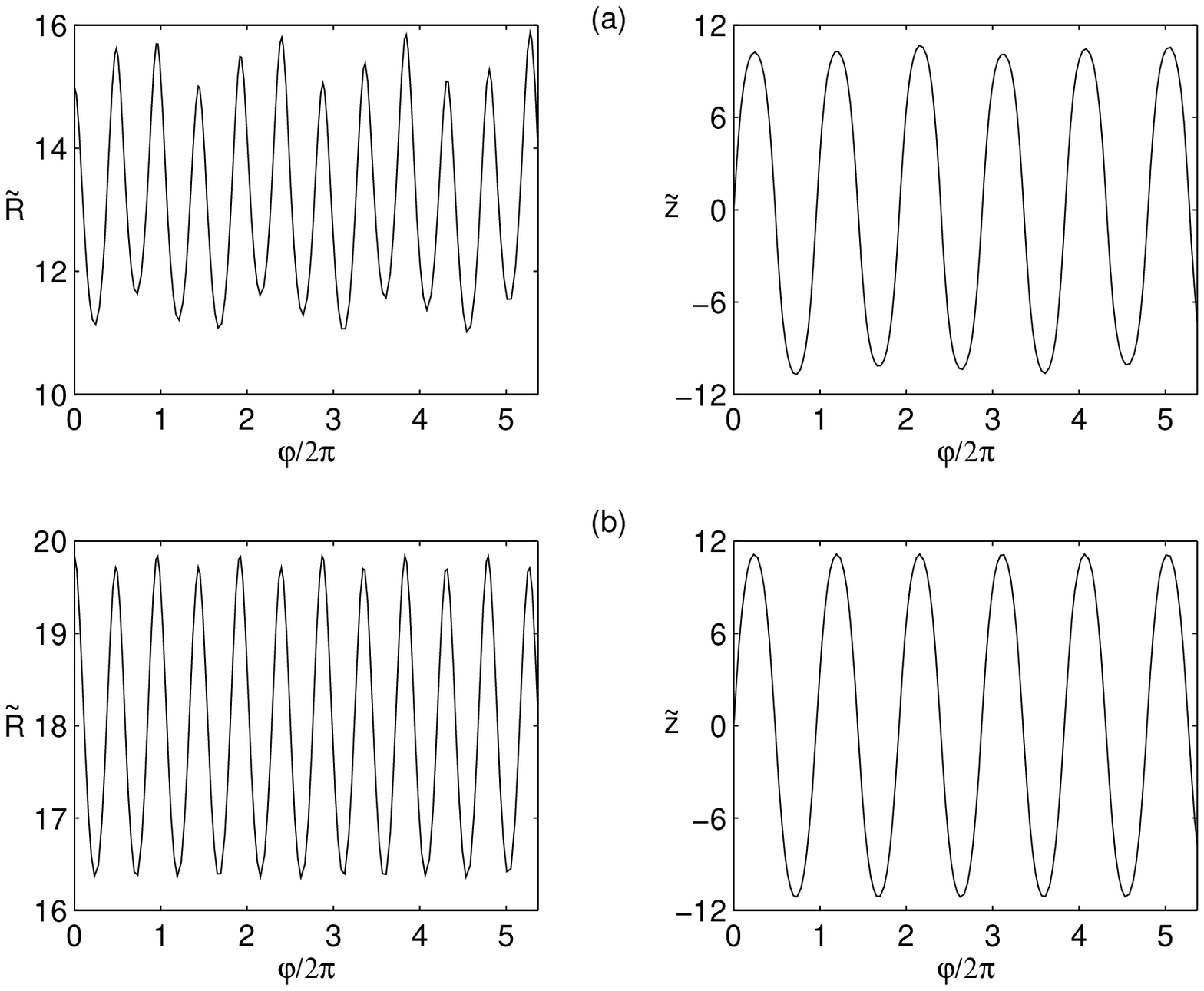}
\caption{The coordinates $\tilde{R}$ and $\tilde{z}$ as functions of
the number of revolutions $\varphi/2\pi$ for the orbits shown in
Fig.\ \ref{fig13}. (a) Geodesic and (b) Newtonian orbit. } \label{fig14}
\end{figure}

\begin{figure}
\centering
\includegraphics[scale=0.5]{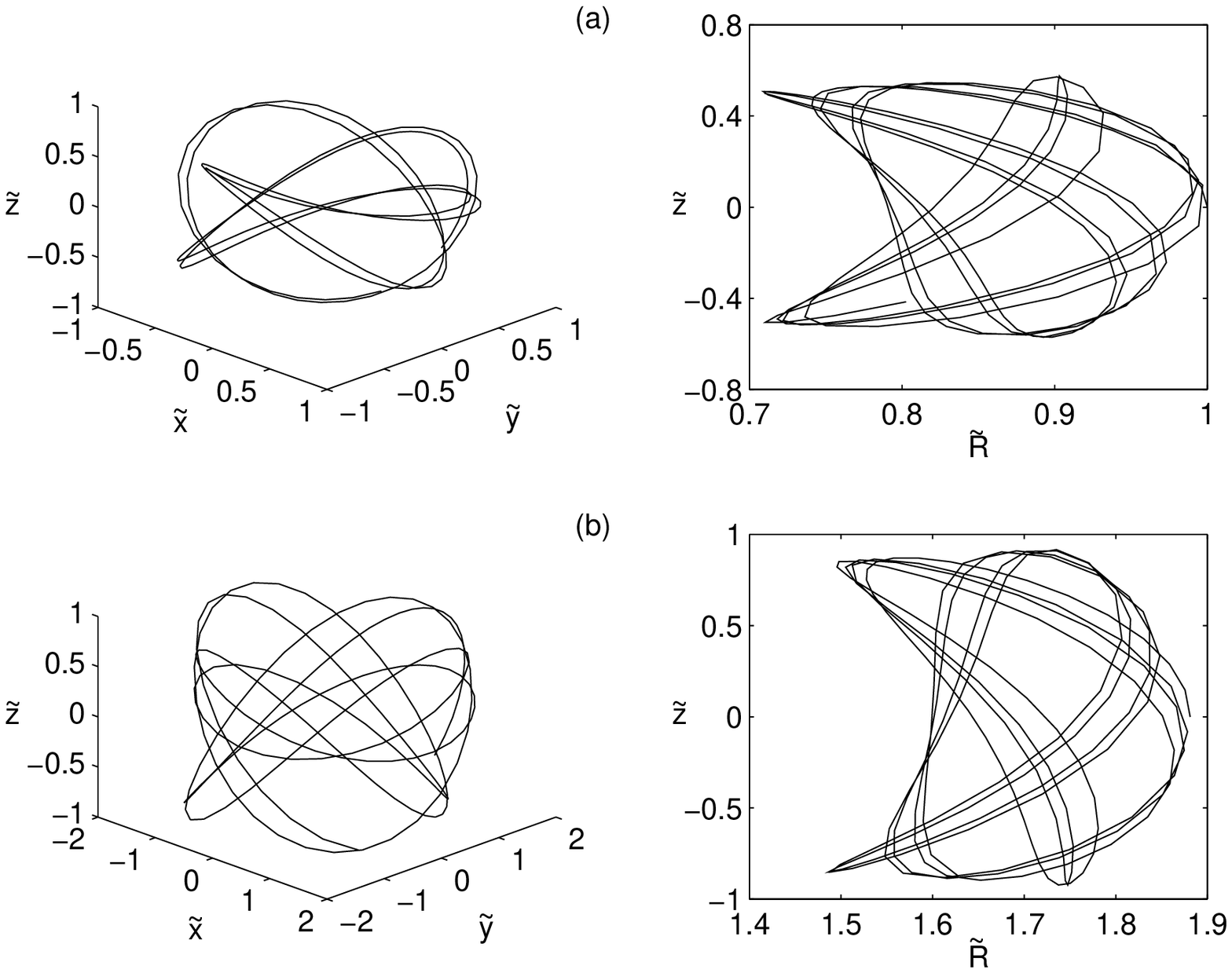}
\caption{(a) Geodesic orbits for the first Miyamoto-Nagai model with
parameters $\tilde{a}=1$, $\tilde{b}=1$. Initial conditions: $\tilde{R}_0=1$,
$\psi=45^o$ with energy $\mathcal{E}\approx 0.196$ and angular momentum $\bar{h} \approx
0.867$. (b) Newtonian orbit
with energy $\mathcal{E}-1$ and same angular momentum. Initial conditions: $\tilde{R}_0 
\approx 1.88$ and $\psi \approx 38.01^o$.} \label{fig15}
\end{figure}

\begin{figure}
\centering
\includegraphics[scale=0.5]{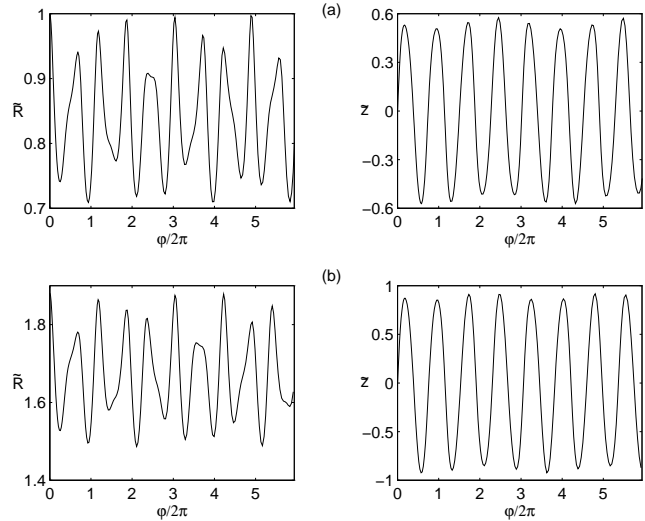}
\caption{The coordinates $\tilde{R}$ and $\tilde{z}$ as functions of
the number of revolutions $\varphi/2\pi$ for the orbits shown in
Fig.\ \ref{fig15}. (a) Geodesic and (b) Newtonian orbit. } \label{fig16}
\end{figure}
In Fig.\ \ref{fig13}--\ref{fig16} we compare both orbits for the first Miyamoto-Nagai model 
with parameters $\tilde{a}=1$ and $\tilde{b}=1$. For the orbits shown in Fig.\ \ref{fig13}--\ref{fig14}
large radii have been chosen as initial conditions. As expected, General Relativistic 
effects are small. A noticeable difference is seen in a periodic variation of the
amplitude of the $\tilde{R}$ coordinate as function of the azimuthal angle in the
relativistic case. On the other hand, the orbits shown in Fig.\ \ref{fig15}--\ref{fig16}
have been computed near the center where the gravitational fields are strong, and 
the deviation between relativistic and Newtonian orbits are much more pronounced.

\section{First Order Effects of Galactic Rotation on the Rotation Profiles} \label{sec_rot}

The galactic models presented in the last sections are all static. Astrophysical objects are 
expected to have angular momentum, thus more realistic models should incorporate 
rotation. An exact stationary General Relativistic galactic model could be constructed, for 
example, if one managed to find a closed form of the Kerr metric in terms of elementary functions in stationary isotropic coordinates,
and then repeat the procedure outlined in the previous sections to calculate the matter properties. 
Although this cannot be done for the \emph{exact} Kerr solution, the same is not true for
the first order expansion of the Kerr metric with respect to the rotation parameter. It reads
\begin{equation} \label{eq_metr_kerr}
\mathrm{d}s^2=\left( 1-\frac{2GM}{rc^2}\right)c^2\mathrm{d}t^2-\frac{\mathrm{d}r^2}
{1-\frac{2GM}{rc^2}}-r^2(\mathrm{d}\theta ^2+\sin^2 \theta \mathrm{d}\varphi^2) 
+\frac{4GM\alpha}{rc} \sin^2 \theta \mathrm{d}t \mathrm{d} \varphi \mbox{,}
\end{equation}
where $\alpha$ is the rotation parameter. If we transform metric Eq.\ (\ref{eq_metr_kerr}) to cylindrical isotropic coordinates 

$(t,R,z,\varphi)$ and express it in terms of the function $f$ given by Eq.\ (\ref{eq_iso_schw}) 
we get
\begin{equation} \label{eq_metr_kerr_iso}
\mathrm{d}s^2=\left( \frac{1-f}{1+f}\right)^2 c^2\mathrm{d}t^2-(1+f)^4\left( \mathrm{d}R^2+
\mathrm{d}z^2+R^2\mathrm{d}\varphi ^2 \right) 
+\frac{32\alpha R^2f^3c^5}{G^2M^2(1+f)^2}\mathrm{d}t \mathrm{d} \varphi \mbox{.}
\end{equation}
Now the function $f$ may be replaced by any of the forms presented in the previous 
sections to generate spacetimes with matter. However, a direct 
calculation using Einstein's field equations and metric Eq.\ (\ref{eq_metr_kerr_iso}) 
shows that the influence of rotation on the physical variables of the matter is of 
at least $\mathcal{O}(\alpha^2)$, and thus to have a consistent correction one should also start
with metric form Eq.\ (\ref{eq_metr_kerr_iso}) corrected at least to $\mathcal{O}(\alpha^2)$, but
it does not seem possible to express this particular form of the metric with second 
order corrections of the rotation parameter. Despite this, a consistent first order 
correction of the rotation profiles can be calculated. 

We rewrite Eq.\ (\ref{eq_metr_kerr_iso}) as
\begin{equation} \label{eq_metr_kerr_gen}
\mathrm{d}\tau^2=g_{tt}\mathrm{d}t^2-g_{RR}\left( \mathrm{d}R^2+
\mathrm{d}z^2+R^2\mathrm{d}\varphi ^2 \right)+2g_{t\varphi}
\mathrm{d}t \mathrm{d} \varphi \mbox{,}
\end{equation}
where for convenience we have set $c=1$. Let $v^{a}=(\dot{t},0,0,\dot{\varphi})$ be the 
velocity four-vector, where dots denote differentiation with respect to $\tau$. The geodesic equation for circular orbits is
\begin{equation}
g_{tt,R}\dot{t}^2+2g_{t\varphi ,R}\dot{t}\dot{\varphi}-(R^2g_{RR})_{,R}\dot{\varphi}^2=0 
\Rightarrow \frac{\dot{\varphi}}{\dot{t}}=\frac{ g_{t\varphi ,R}\pm \sqrt{ (g_{t\varphi ,R})^2+ 
g_{tt,R}(R^2g_{RR})_{,R}}}{(R^2g_{RR})_{,R}} \mbox{.}  \label{eq_geo_rot}
\end{equation}
 Thus the prograde and retrograde angular velocities are in general different. We define an orthonormal 
tetrad $e_{(b)i}$ as
\begin{align}
e_{(t)i}&= \left( \frac{\sqrt{g_{t\varphi}^2+R^2g_{tt}g_{RR}}}{R\sqrt{g_{RR}}},0,0,0 \right) \mbox{,} \notag \\
e_{(R)i} &=(0,-\sqrt{g_{RR}},0,0), \quad  e_{(z)i} = (0,0,-\sqrt{g_{RR}},0) \mbox{,} \notag \\
e_{(\varphi)i} &=\left( \frac{g_{t\varphi}}{R\sqrt{g_{RR}}},0,0,-R\sqrt{g_{RR}} \right) \mbox{,}
\end{align}
and project the components of $v^{a}$ on the tetrad: $\mathsf{v}^{(a)}=e^{(a)}_iv^i$=
$\eta^{(a)(b)}e_{(b)i}v^i$
\begin{subequations}
\begin{align}
\mathsf{v}^{(t)} &= \frac{\sqrt{g_{t\varphi}^2+R^2g_{tt}g_{RR}}}{R\sqrt{g_{RR}}}\dot{t} 
\mbox{,} \label{eq_vt} \\
\mathsf{v}^{(\varphi)} &=-\frac{g_{t\varphi}}{R\sqrt{g_{RR}}}\dot{t}+R\sqrt{g_{RR}}
\dot{\varphi} \mbox{.} \label{eq_vfi}
\end{align}
\end{subequations}
Using Eq.\ (\ref{eq_geo_rot}) in Eq.\ (\ref{eq_vt})--(\ref{eq_vfi}) we get
\begin{equation}
\frac{\mathsf{v}^{(\varphi)}}{\mathsf{v}^{(t)}}=\frac{1}{\sqrt{g_{t\varphi}^2+R^2g_{tt}g_{RR}}}
\left[ -g_{t\varphi}+\frac{R^2g_{RR}}{(R^2g_{RR})_{,R}}\left( g_{t\varphi,R} 
 \pm \sqrt{ (g_{t\varphi,R})^2+g_{tt,R}(R^2g_{RR})_{,R}} \right) \right] \mbox{.} \label{eq_vc_rot_a}
\end{equation}
The terms $g_{t\varphi}^2$ and $(g_{t\varphi,R})^2$ in Eq.\ (\ref{eq_vc_rot_a}) are of 
$\mathcal{O}(\alpha^2)$ and thus can be neglected. Using metric 
Eq.\ (\ref{eq_metr_kerr_iso}), Eq.\ (\ref{eq_vc_rot_a}) can be
rearranged as 
\begin{equation} \label{eq_vc_rot_final}
v_c=c\sqrt{\frac{-2Rf_{,R}}{( 1-f )( 1+f+2Rf_{,R})}}\left[ \pm 1+\frac{24\alpha f^2Rc^4}
{G^2M^2(1+f)^3} \sqrt{\frac{-Rf_{,R}(1-f)}{2(1+f+2Rf_{,R})}} \right] \mbox{,}
\end{equation}
where
 $c$ has been reintroduced. The term outside brackets is the tangential velocity 
Eq.\ (\ref{eq_vcf}). The term with $\alpha$ is always positive, thus the first order 
effect of rotation increases (decreases) the prograde (retrograde) tangential velocity. 

Eq.\ (\ref{eq_vc_rot_final}) calculated with the functions $f$ used in Sec.\ \ref{sec_miy}--\ref{sec_satoh} 
results in
\begin{align}
\frac{v_c}{c} &=\frac{\sqrt{2}\tilde{R}(\tilde{R}^2+\xi)^{1/4}}{[-1+(\tilde{R}^2+\xi)^{1/2}]^{1/2}
[-\tilde{R}^2+\xi+(\tilde{R}^2+\xi)^{3/2}]^{1/2}}\left[ \pm 1+\frac{3\tilde{\alpha}\tilde{R}^2
(\tilde{R}^2+\xi)^{1/4}}{[1+(\tilde{R}^2+\xi)^{1/2}]^3}\sqrt{\frac{2[-1+(\tilde{R}^2+\xi)^{1/2}]}
{-\tilde{R}^2+\xi+(\tilde{R}^2+\xi)^{3/2}}} \right] \mbox{,} \label{eq_vc_rot1} \\
\frac{v_c}{c} &=\frac{\sqrt{2}\tilde{R}\zeta^{3/4}[\tilde{R}^2+(\tilde{a}+\tilde{b})(4\tilde{a}+
\tilde{b})]^{1/2}}{\left[ \zeta^{3/2}-\tilde{R}^2-(\tilde{a}+\tilde{b})(2\tilde{a}+\tilde{b})\right]^{1/2}
\left[ \zeta^{5/2} -\tilde{R}^4-5\tilde{a}(\tilde{a}+\tilde{b})\tilde{R}^2 +(\tilde{a}+\tilde{b})^3
(2\tilde{a}+\tilde{b})\right]^{1/2}} \times \notag \\
& \left[ \pm 1+\frac{3\tilde{\alpha}\tilde{R}^2 \zeta^{3/4}
[\tilde{R}^2+(\tilde{a}+\tilde{b})(2\tilde{a}+\tilde{b})]^2}{[\zeta^{3/2}+\tilde{R}^2
+(\tilde{a}+\tilde{b})(2\tilde{a}+\tilde{b})]^3}\sqrt{\frac{2[\tilde{R}^2+(\tilde{a}+
\tilde{b})(4\tilde{a}+\tilde{b})][\zeta^{3/2}-\tilde{R}^2-(\tilde{a}+\tilde{b})(2\tilde{a}+
\tilde{b})]}{\zeta^{5/2}-\tilde{R}^4-5\tilde{a}(\tilde{a}+\tilde{b})\tilde{R}^2 +(\tilde{a}+
\tilde{b})^3(2\tilde{a}+\tilde{b})}} \right] \mbox{,} \label{eq_vc_rot2}
\end{align}
where in Eq.\ (\ref{eq_vc_rot1}) $\xi=(\tilde{a}+\tilde{b})^2$ for the first Miyamoto-Nagai 
model (Sec.\ \ref{sub_sec_miy1}), $\xi=\tilde{a}(\tilde{a}+2\tilde{b})$ for the Satoh model 
(Sec.\ \ref{sec_satoh}), and in Eq.\ (\ref{eq_vc_rot2}) $\zeta=\tilde{R}^2+(\tilde{a}+\tilde{b})^2$ 
for the second Miyamoto-Nagai model (Sec.\ \ref{sub_sec_miy2}). The dimensionless variables 
were defined in Sec.\ \ref{sub_sec_miy1} and $\tilde{\alpha}=\alpha/r_s$.

\begin{figure}
\centering
\includegraphics[scale=0.68]{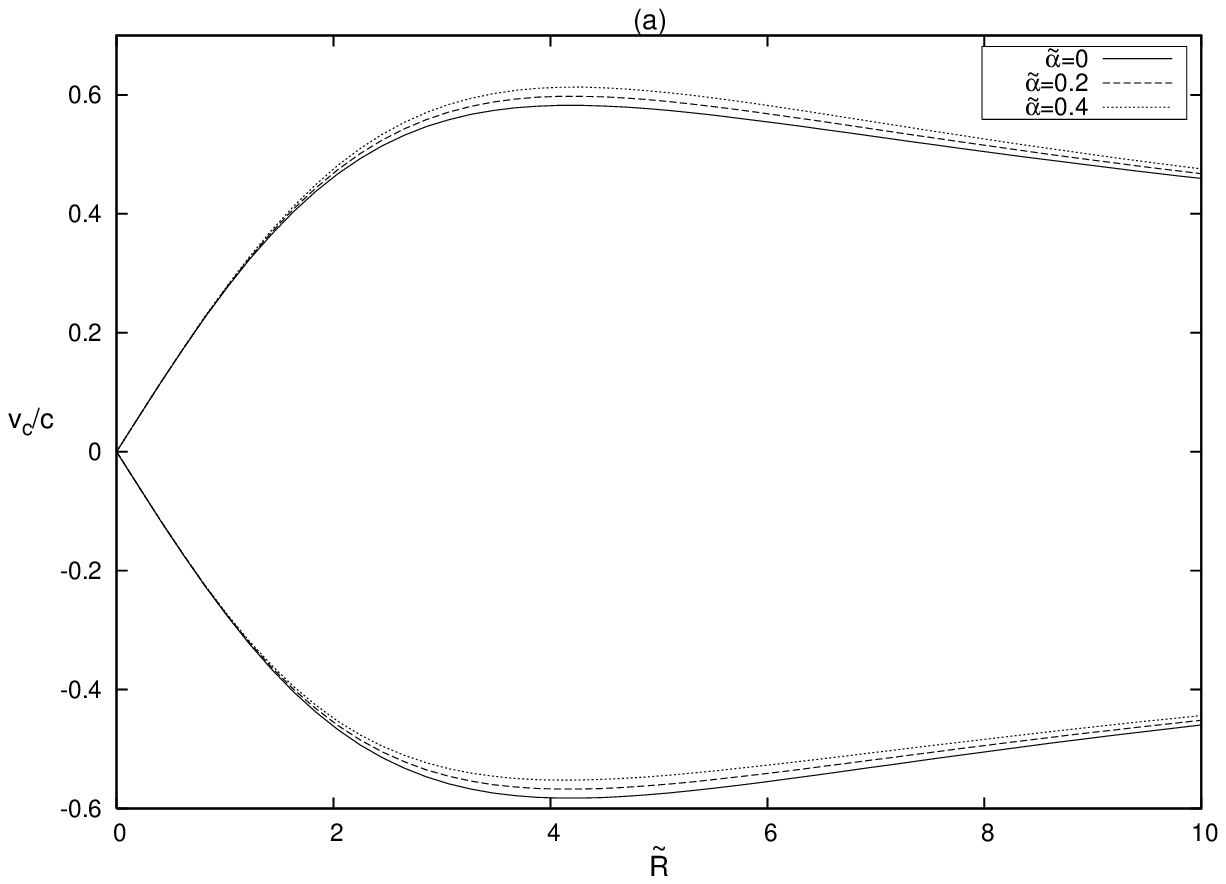}\\
\includegraphics[scale=0.68]{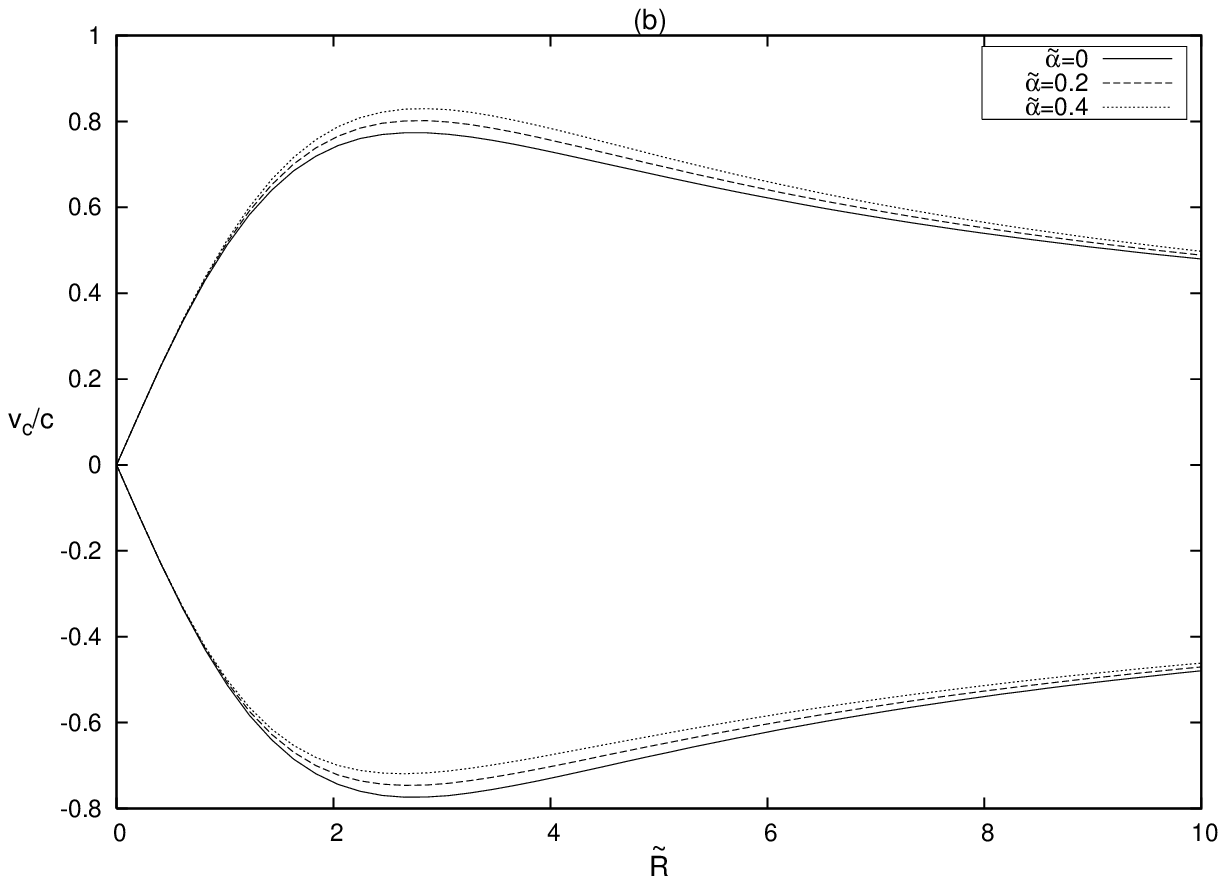}
\caption{Prograde (upper) curves and retrogade (lower) curves of tangential velocity Eq.\ (\ref{eq_vc_rot1}) for the
first Miyamoto-Nagai model with parameters (a) $\tilde{a}=1$, $\tilde{b}=2$ and
(b) $\tilde{a}=1$, $\tilde{b}=1$. Rotation parameter: $\tilde{\alpha}=0$ (solid line),
$\tilde{\alpha}=0.2$ and $\tilde{\alpha}=0.4$.} \label{fig17}
\end{figure}

\begin{figure}
\centering
\includegraphics[scale=0.68]{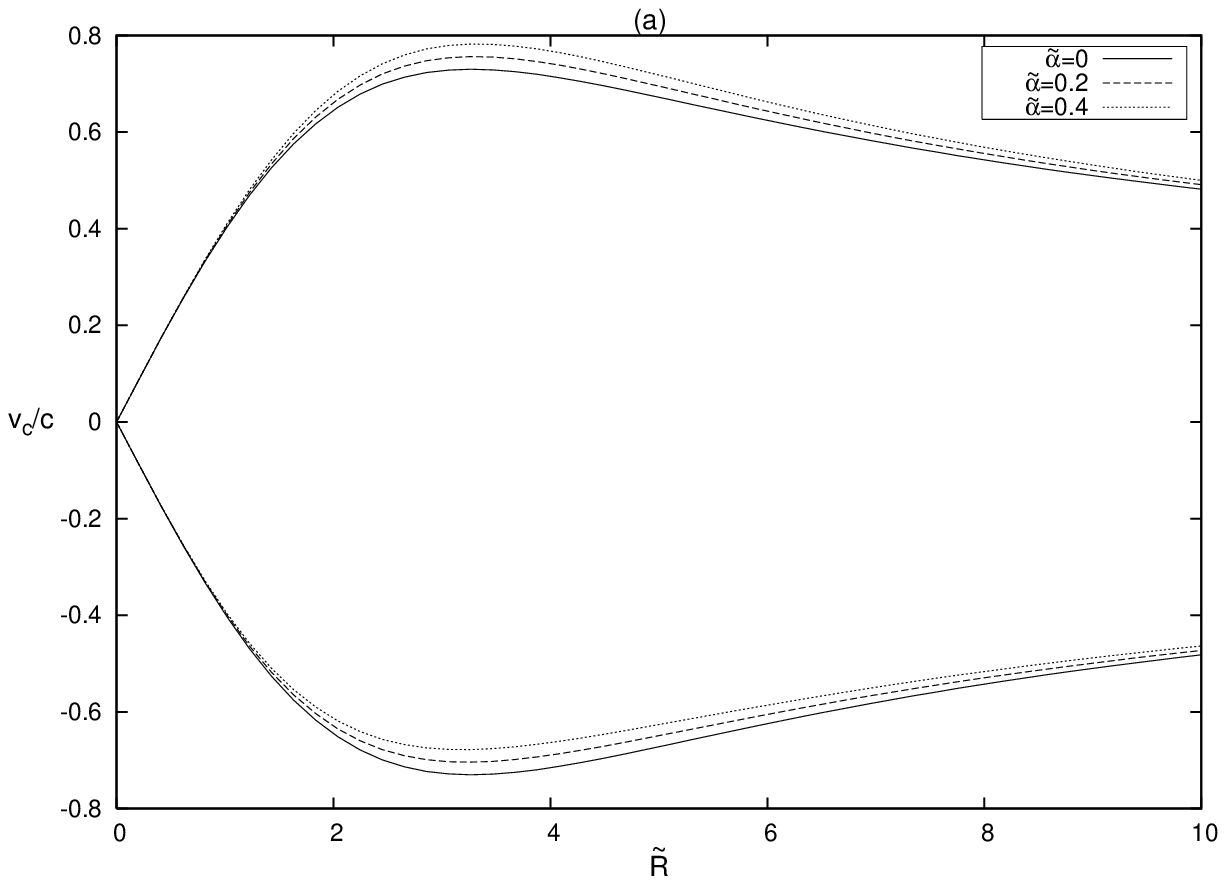}\\
\includegraphics[scale=0.68]{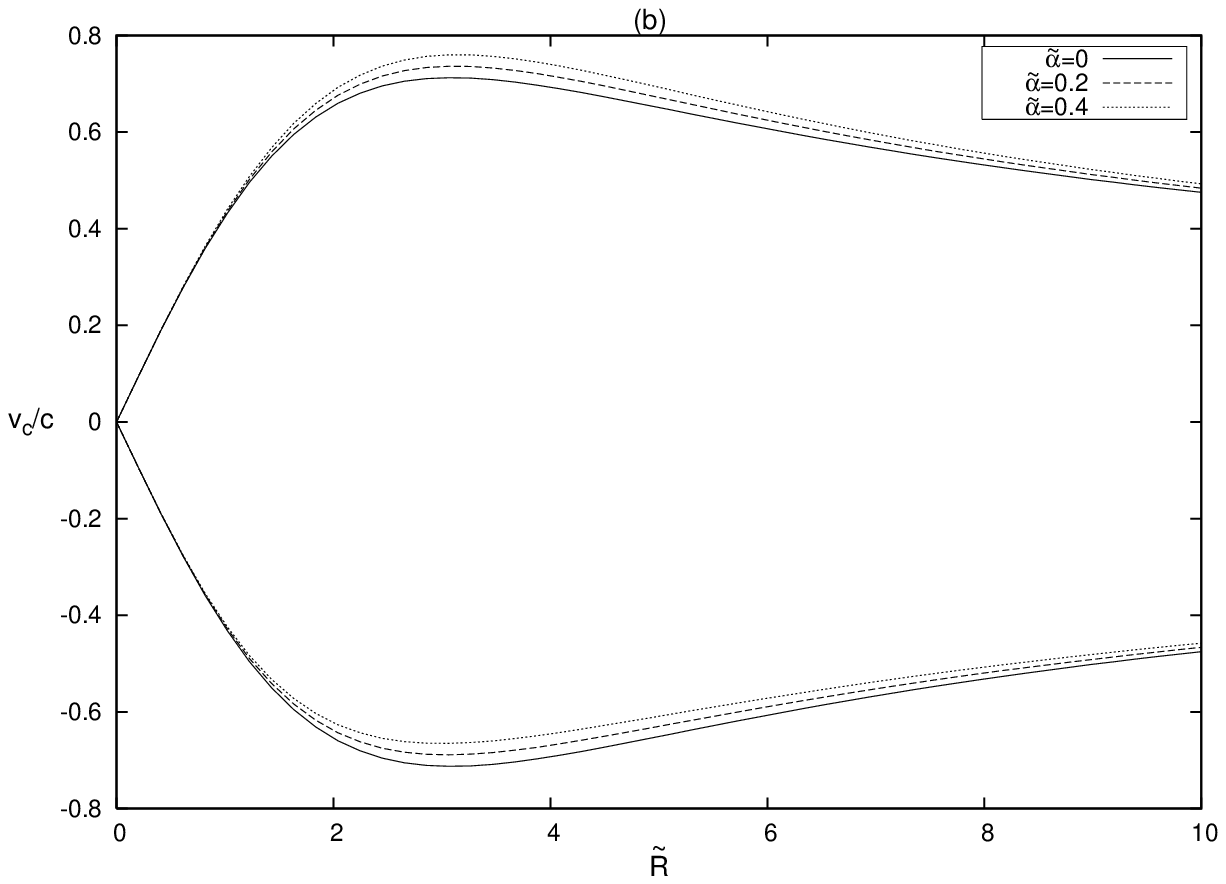}
\caption{Prograde (upper) curves and retrogade (lower) curves of tangential velocity for
(a) the second Miyamoto-Nagai model Eq.\ (\ref{eq_vc_rot2}) with parameters $\tilde{a}=1$, 
$\tilde{b}=2$ and (b) for the Satoh model Eq.\ (\ref{eq_vc_rot1}) with parameters $\tilde{a}=1$,
$\tilde{b}=2$. Rotation parameter: $\tilde{\alpha}=0$ (solid line), $\tilde{\alpha}=0.2$ and $\tilde{\alpha}=0.4$.} \label{fig18}
\end{figure}

In Fig.\ \ref{fig17}(a)--(b) we plot the curves of progade and retrogade tangential velocity
for the first Miyamoto-Nagai model with parameters $\tilde{a}=1$, $\tilde{b}=2$ and
$\tilde{a}=1$, $\tilde{b}=1$ with rotation parameter $\tilde{\alpha}=0$, $\tilde{\alpha}=0.2$ 
and $\tilde{\alpha}=0.4$. The same curves for the second Miyamoto-Nagai model
and for the Satoh model with parameters $\tilde{a}=1$, $\tilde{b}=2$ are 
shown in Fig.\ \ref{fig18}(a)--(b).
\section{Discussion} \label{sec_discuss}

Using a special form for the metric in cylindrical isotropic coordinates, we have constructed
what may be interpreted as the General Relativistic versions of some known Newtonian
models for three-dimensional distributions of matter in galaxies, like those formulated by
Miyamoto and Nagai (\citealt{b5}; \citealt{b6}) and \citet{b7}. For the first Miyamoto- 
Nagai potential and for one Satoh potential we obtained matter distributions that have similar 
properties: their energy-momentum tensor is diagonal with isotropy of pressure in the radial 
and azimuthal directions; energy density and the ``effective Newtonian density'' are positive everywhere as 
well as the stresses (pressures) in the
vertical direction. For particular values of the free parameters we found that the dominant energy 
condition is also satisfied. Stability analysis of the circular geodesic orbits on the galactic plane 
using an extension of the Rayleigh criteria of stability shows stable orbits for some combination 
of free parameters. The second Miyamoto-Nagai potential function, after diagonalization of the 
energy-momentum tensor, yield matter distributions with positive energy density and three different 
pressures. Also in this case, free parameters can be chosen so that all energy 
conditions are satisfied and circular orbits are stable on the galactic plane.

As an example of application of the models, we have numerically calculated some geodesic orbits 
for one of the potentials and compared them with the Newtonian orbits with the same 
energy and angular momentum. Near the central regions where the gravitational fields are strong, 
the motion of particles is considerably altered by General Relativistic effects.  

 We also calculated the first order effects of galactic rotation on the tangential velocity of circular orbits 
on the galactic plane using an approximate form of the Kerr metric expressed in cylindrical isotropic 
coordinates. In general, rotation increases the progade tangential velocity and has an opposite
effect on the retrogade tangential velocity. 

It should be mentioned that the stability study of the models presented based on the extension of Rayleigh 
criteria of stability
is very limited. A more realistic stability analysis should rely on the first order perturbed 
General Relativistic fluid equations taking into account two spatial coordinates, which may be 
not a trivial task (see, for instance, \citet{b29} for the one dimensional case). This will be subject of further investigation.

\section*{Acknowledgments}
D.\ Vogt thanks CAPES for financial support. P.\ S.\ Letelier thanks CNPq and FAPESP for financial support. This research has made use of NASA's Astrophysics Data System.

\end{document}